\documentclass[prx, twocolumn, superscriptaddress, longbibliography, showpacs, nofootinbib]{revtex4-1}

\usepackage{fancyhdr}
\fancyheadoffset[L,R]{\headrulewidth}
\renewcommand{\headrulewidth}{0.6pt}

\usepackage{cancel}
\usepackage[normalem]{ulem}
\usepackage{graphicx}
\usepackage{tikz,pgf}
\usepackage{color}
\usepackage{array}
\usepackage{xcolor}
\usepackage[dvips]{epsfig}
\usepackage{epsfig}
\usepackage{enumerate}
\usepackage[latin1]{inputenc}
\usepackage[T1]{fontenc}
\usepackage[hang]{subfigure}
\usepackage{bbm, dsfont}
\usepackage{amsfonts,amsmath,amssymb,stmaryrd}
\usepackage{ulem}
\usepackage{mathrsfs}
\usepackage{bbold}
\usepackage{simplewick}

\definecolor{linkcolor}{rgb}{0,0,0.6} 

\newcommand{\qq}{\begin{eqnarray}}
\newcommand{\qqq}{\end{eqnarray}}

\newcommand{\p}{\partial}

\newcommand{\bfr}{\mathbf{r}}

\newcommand{\bfu}{{\bf u}}
\newcommand{\bfJ}{{\bf J}}

\begin{document}

\title{Cluster phases and bubbly phase separation in active fluids: \\
Reversal of the Ostwald process
}

\author{Elsen Tjhung} 
\affiliation{DAMTP, Centre for Mathematical Sciences, University of Cambridge, Wilberforce Road, Cambridge CB3 0WA, UK}

\author{Cesare Nardini} 
\affiliation{Service de Physique de l'\'Etat Condens\'e, CNRS UMR 3680, CEA-Saclay, 91191 Gif-sur-Yvette, France}

\author{Michael E. Cates} 
\affiliation{DAMTP, Centre for Mathematical Sciences, University of Cambridge, Wilberforce Road, Cambridge CB3 0WA, UK}


\begin{abstract}
It is known that purely repulsive self-propelled colloids can undergo bulk liquid-vapor phase separation. 
 In experiments and large scale simulations, however, more complex steady states are also seen, comprising a dynamic population of dense clusters in a sea of vapor, or dilute bubbles in a liquid. Here we show that these microphase-separated states should emerge {\em generically} in active matter, without any need to invoke system-specific details. We give a coarse-grained description of them, and predict transitions between regimes of bulk phase separation and microphase separation. We achieve these results by extending the $\phi^4$ field theory of passive phase separation to allow for all local currents that break detailed balance at leading order in the gradient expansion.
These local active currents, whose form we show to emerge from coarse-graining of microscopic models, include a mixture of irrotational and rotational contributions, and can be viewed as arising from an effective {\em nonlocal} chemical potential. 
 Such contributions influence, and in some parameter ranges reverse, the classical Ostwald process that would normally drive bulk phase separation to completion.
\end{abstract}


\maketitle


Active colloidal fluids are non-equilibrium systems in which individual particles continually consume fuel in order to self-propel~\cite{ramaswamy2017active,Marchetti2013RMP}. 
Time-reversal symmetry is broken locally, leading to features impossible in thermal equilibrium.
One of these is motility-induced phase separation (MIPS) where an assembly of repulsive but active particles phase separates into bulk dense and dilute regions~\cite{Tailleur:08,Cates:15,Fily:12}. 
The MIPS phenomenology has been confirmed in simulations~\cite{Fily:12,baskaranABP,stenhammar2014phase,baskaranABPPRE,Brader:15} and 
explored experimentally~\cite{Speck:13,liu2017motility}.
Although a far-from-equilibrium phenomenon, 
MIPS was first understood via an approximate mapping onto an effective equilibrium system undergoing standard liquid-vapor separation~\cite{Tailleur:08,Solon:15c,Cates:15}.  
This has led to speculation that for MIPS, time reversal symmetry is effectively restored at the macroscopic level in steady state \cite{Speck2014PRL,Tailleur:08,fodor2016far,Brader:15,Maggi:15,szamel2016theory,nardini2017entropy}. 

\begin{figure*}
\begin{centering}
\includegraphics[width=0.8\textwidth]{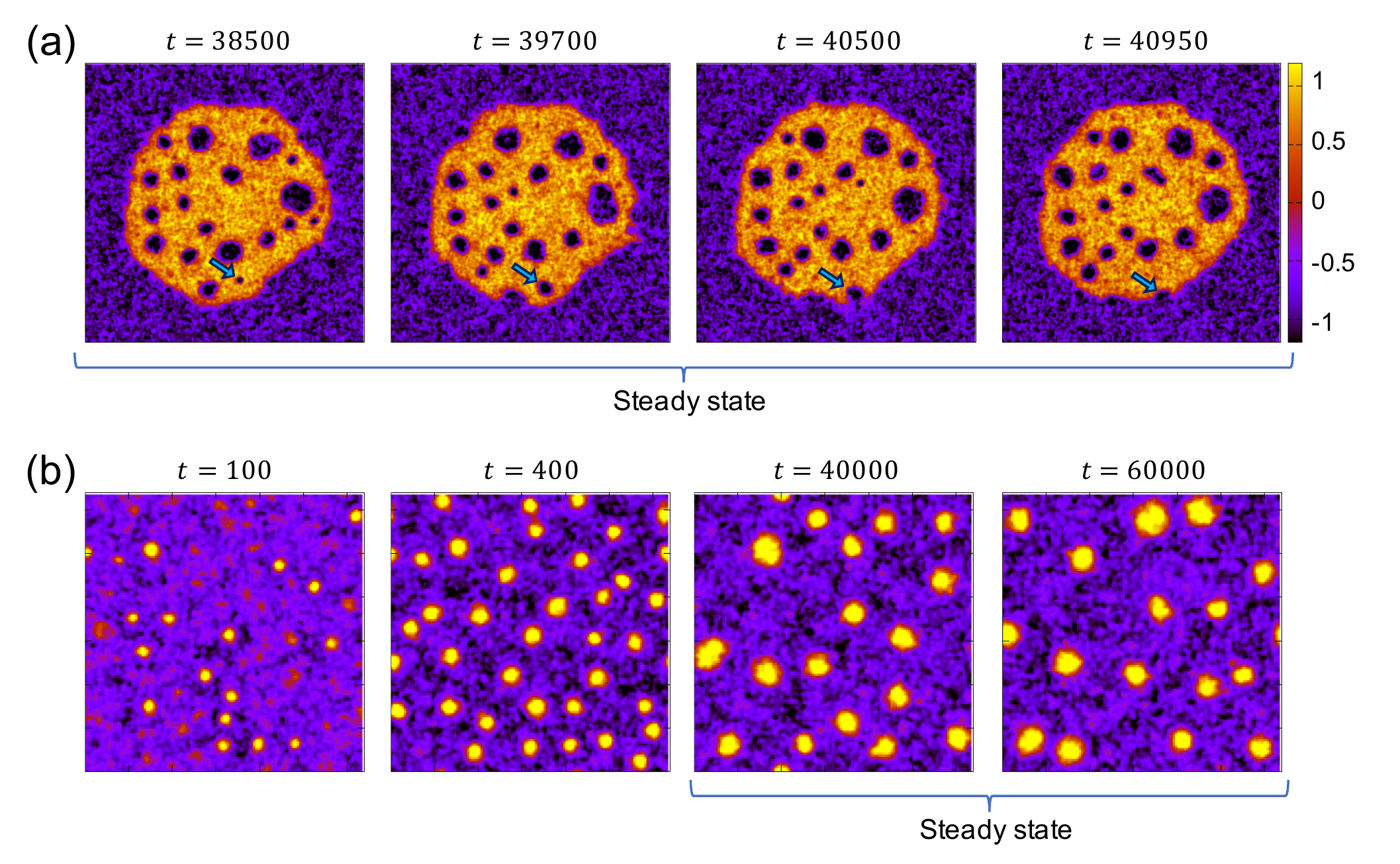}
\par\end{centering}
\caption{
(a) Bubbly phase separation as described in our field theory, a phase coexistence of boiling liquid (yellow) with a vapor phase (blue).
Vapor bubbles are continuously created inside the bulk liquid phase, which are then expelled to the exterior vapor phase (indicated by blue arrow in the figure, also see Suppl. Movie 4).
The vapor phase can then diffuse back inside the liquid and the process continues indefinitely. Time reversal symmetry is manifestly broken in this steady state.
(b) Arrested phase separation: initially, the system phase separates into dense phase and dilute phase.
However, the average cluster size saturates to a finite value at steady state. Additional steady states, representing (a) a cluster phase in coexistence with excess dense liquid and (b) a strict bubble phase are found by inverting the order parameter scale shown at top right. This exploits invariance of the model under the duality mapping explained in Section \ref{model}.}
\label{fig:snapshots}
\end{figure*}

However, simulations of repulsive active Brownian particles (ABPs) reveal that during coarsening, 
the macroscopic droplets of dense liquid arising via MIPS host a population of mesoscopic vapor bubbles that are continuously created in the bulk, coarsen, and are ejected into the exterior vapor~\cite{stenhammar2014phase}.
This suggests that the steady state involves coexistence between a boiling liquid and excess vapor which we call {\it bubbly phase separation}. 
The life-cycle of the mesoscopic bubbles, if persistent even when coarsening is completed, would confirm that time reversal symmetry remains manifestly broken in this steady state. While not fully established for ABPs simulations, we present results below for continuum models that definitely show this behavior. The `boiling liquid' is itself a microphase-separated state of vapor bubbles (or `bubble phase') surrounded by dense liquid.  

Puzzlingly, these observations on active Brownian particles are almost dual to the experimental findings on synthetic self-propelled particles~\cite{Palacci:12,Speck:13,thutupalli2017boundaries}, 
which can instead form dense living clusters. In these experiments, self-propelled colloids indeed aggregate but the clusters show highly dynamic particle exchange and, importantly, they never coarsen to reach the system size but saturate at a mesoscopic scale. (Such clusters can also show crystalline order, which we ignore here.) The `boiling liquid' referred to above is akin to a cluster phase in which the dense and dilute regions have been interchanged. 
In the literature, several mechanisms for the emergence of cluster phases have been suggested, based on long range interactions  
(whether hydrodynamic \cite{tiribocchi2015active,matas2014hydrodynamic,thutupalli2017boundaries} or chemotactic \cite{liebchen2015clustering,saha2014clusters}), 
or on other, system-specific, effects \cite{Brader:15,alarcon2017morphology,prymidis2015self,mani2015effect,mognetti2013living}. 
Cluster phases are also found in models of phase separation coupled to birth-and-death processes, 
where the number density of one chemical species is not conserved~\cite{CatesPNAS,zwicker2017growth,Zwicker2015}. 

In this paper, we show that the observed occurrence of cluster phases and  of bubbly phase separation do not require system-specific explanations, but already emerge from a completely generic continuum model of active phase separation. This differs from equilibrium phase separation models by allowing for broken time-reversal symmetry (TRS) at mesoscopic scales.  At leading order in an expansion in gradients of the density, our model admits two distinct contributions to the particle current, only one of which was considered in previous studies of active phase separation~\cite{Tailleur:08,Speck:2014:PRL,stenhammar2013continuum,Wittkowski14,solon2018generalized,solon2018generalized2}. 
The additional current term was considered in~\cite{nardini2017entropy}, but its role in phase equilibria not studied there; here we show it to be crucial to predicting microphase separation. The unified description that we present, in which clusters and bubbles are dual to one another (see Section \ref{model}), not only accounts for the existence of cluster phases (either alone or coexisting with an excess dense fluid), and bubbly phase separation, but also predicts the existence of a strict bubble phase in which the whole system is filled by a dense liquid that supports a fluctuating, but not coarsening, population of mesoscopic vapor bubbles. This phase was so far not reported in either experiments or particle-based simulations of active matter. Snapshots of the various steady states predicted by the continuum model considered in this paper, are given in Fig. \ref{fig:snapshots}.

In equilibrium systems subject to particle diffusion without momentum conservation, bulk phase separation is driven by the Ostwald process~\cite{Bray,CatesJFM:2018}. Here, local equilibrium requires the Laplace pressure jump across an interface to increase with its curvature. The chemical potential at  a curved interface is thereby shifted, creating a compositional excess outside each droplet that decreases in magnitude with increasing droplet size. The local excess sets boundary conditions on a quasi-static solution of the diffusion equation, which slowly drives material down the density gradients from smaller to larger droplets. Small droplets evaporate, and large ones grow; the mean droplet size increases forever, ultimately leading to complete phase separation. Exactly similar arguments apply to the growth of vapor bubbles on the opposite side of the phase diagram.

A major achievement of the present work is to show analytically how the Ostwald process can become reversed in active fluids. This unexpected result is supported by numerical findings that demonstrate microphase-separated steady states in the parameter regimes where the reverse Ostwald process is predicted. Interestingly, our model's parameter values can be chosen to stabilize {\em either} vapor bubbles dispersed in a homogeneous dense liquid, {\em or} liquid clusters dispersed in a vapor, but not both at once. This might help explain why computer simulations of hard-core active Brownian particles 
report only bubbles, and not cluster phases ~\cite{Fily:12,baskaranABP,stenhammar2014phase,baskaranABPPRE} with the reverse being true for experiments on autophoretic Janus colloids \cite{Palacci:12,Speck:13,thutupalli2017boundaries}. The microscopic interactions are clearly different in the two systems (and much more complex for the experiments) so perhaps they represent opposing halves of the parameter space of our coarse-grained model. 

Our starting point is Model B, the $\phi^4$ theory of diffusive phase ordering of a conserved order parameter $\phi$ with time-reversal symmetry~\cite{hohenberg1977theory}. Crucially though, and differently to previous attempts in this direction~\cite{Tailleur:08,Speck:2014:PRL,stenhammar2013continuum,Wittkowski14,solon2018generalized,solon2018generalized2}, 
we consider the field theory obtained by adding {\em all} terms that break time-reversal symmetry to leading order in $\nabla$ and $\phi$, while retaining mass conservation, $\phi$-independent mobility, and hence additive noise. (The latter is a crucial simplification for analytic work.)
Following~\cite{nardiniperturbative2016}, we call the ensuing theory Active Model B+ (AMB+).

Like Model B \cite{hohenberg1977theory}, AMB+ is naturally introduced on the basis of symmetries, conservation laws, and the gradient expansion. However, we will show that a closely related continuum description -- the main difference being that mobility is not constant and hence that noise is multiplicative -- can be obtained from explicit coarse graining of microscopic models of self-propelled particles. This strongly suggests that the TRS-breaking terms in AMB+ are not forbidden by any microscopic mechanism that we might have overlooked, and hence that its prediction of reverse Ostwald regimes are generic, in the same way that Model B is generic for equilibrium binary fluid phase separation \cite{Bray}. Linking microscopic to continuum field-theoretic models also paves the way to future investigations of cluster phases and bubbly phase separation in terms of microscopic parameters, and hence to the possibility of better controlling these new phases of matter experimentally. It might also suggest where to look experimentally for the strict bubble phase mentioned above.

The paper is organised as follows. In Section \ref{model} we introduce AMB+ on the basis of symmetry arguments. In Section \ref{binodals} we analyse the model at mean-field level, looking at stationary phase-separated solutions with spherical symmetry, and deriving coexisting densities as a function of the droplet radius. Such analysis is preliminary to the main analytical results of the paper, derived in Section \ref{Ostwald-ripening}. There, we construct the mean-field phase diagram of AMB+ by
analysing the Ostwald process. 
In Section \ref{noise}, we finally switch on the noise in the field theory and, by means of direct numerical simulations, we show that microphase separations arise in direct correspondence to the regions where Ostwald process is reversed. 
Statistical properties of the microphase separated states are also analysed.
In Section \ref{explicit-coarse-graining}, we
show that the active currents responsible of microphase separation in AMB+ can be derived from explicit coarse-graining of a model of self-propelled particles. We conclude in Section \ref{conclusion} with a discussion of future perspectives. Appendices contain details for those readers who are interested in the technical aspects of this work.

\section{Active Model B+} \label{model}

An equilibrium system undergoing diffusive fluid-fluid phase separation is governed on continuum scales by Model B in the classification of Hohenberg and Halperin~\cite{hohenberg1977theory}. Model B describes the dynamics of a scalar field $\phi(\bfr,t)$ which, for the liquid-vapor transition, is a linear transform of the particle density $\rho(\bfr,t)$, chosen such that $\phi = 0$ at the mean-field critical point density $\rho_c$ of the model \cite{chaikin2000principles}.
The dynamics of $\phi$ follows a mass conservation law:
\qq
\p_t\phi &=& -\nabla \cdot \left( \bfJ+\sqrt{2D M[\phi]}\mathbf{\Lambda} \right) \,,  \label{eq:MB}\\
\bfJ       &=& -M[\phi]\nabla\mu_{eq} = -M[\phi]\nabla\frac{\delta F}{\delta \phi} \,,  \label{eq:MBJ}
\qqq
where $\mathbf{\Lambda}(\mathbf{r},t)$ is a Gaussian white noise with zero mean and unit variance. 
$M[\phi]$, which is positive definite, is the mobility and $D$ is the temperature. Model B has an equilibrium structure, meaning that the deterministic current $\bfJ$ is proportional to the negative gradient of an equilibrium chemical potential $\mu_{eq}$,
which is itself derived from a free energy functional, $\mu_{eq} ={\delta \mathcal{F}}/{\delta\mathcal{\phi}}$. This is taken of square-gradient, $\phi^4$ form:
\begin{equation}
\mathcal{F}[\phi] = \int \bigg\{ \underbrace{\frac{a}{2}\phi^2 + \frac{b}{4}\phi^4}_{f(\phi)} + \frac{K}{2}|\nabla\phi|^2 \bigg\} d\mathbf{r} \,, \label{eq:F}
\end{equation}
where $f(\phi)$ is the bulk free energy density and $K>0$. By choosing $\phi$ to vanish at $\rho = \rho_c$, we have set $f''(0) = f'''(0) = 0$ so that there is no cubic term in $f(\phi)$. This is sufficient reason to work with $\phi$ rather than the particle density $\rho$. Note also that any linear term in $f(\phi)$ would contribute a constant to $\mu_{eq}$ and therefore have no effect. Thus the local part of the free energy $f$ is even to quartic order as written, with the critical point at $a = 0$ and phase separation for $a<0$. (These statements hold without loss of generality, even if the microscopic interactions are asymmetric.) The neglect of terms in $f$ beyond quartic, which would arise for example from expanding the ideal gas entropy term $\rho\ln\rho$, is strictly justified only when the order parameter $\phi$ remains small. The coexisting mean-field binodal densities $\pm\phi_b$ obey $\phi_b = \sqrt{-a/b}$; these should therefore map under the linear transform onto $\rho = \rho_c\pm\Delta\rho_b$ with $\Delta\rho_b \ll\rho_c$. 

We assume this holds here, but also assume that the binodals lie outside the Ginzburg interval \cite{chaikin2000principles} so that we can avoid consideration of critical phenomena. (These are the topic of a separate paper~\cite{caballeroBplus}.) In the language of Bray \cite{Bray}, we are thus interested in phase separation in the neighborhood of the zero-temperature dynamical fixed point, where noise terms might be significant for kinetics (for instance in allowing nucleation) but do not dominate the steady-state statistics, at least in the equilibrium case of Model B. The universality of this fixed point means that qualitative predictions, such as power-law scalings in time for the mean droplet size ($R\sim t^{1/3})$, remain valid even beyond the regime of small $\phi$ to which the model at first appears restricted ~\cite{Bray,chaikin2000principles}.

With the above  proviso of avoiding the critical region, we can now rescale $\phi$ to set the bulk binodals of Model B at $\pm\phi_b = \pm 1$. Alternatively we may achieve this by choosing
\qq -a =  b = A>0
\qqq
which will apply from now on. The equilibrium interfacial tension then obeys $\sigma_{eq} = (8KA/9)^{1/2}$, and the interfacial width $\xi_{eq} = (K/2A)^{1/2}$ \cite{Bray}. 

Finally, we note that a mobility functional $M[\phi]$ appears both as a factor in the deterministic current $\bfJ$ and as a multiplier of the noise $\mathbf{\Lambda}$. This dual role is required by TRS, but it also emerges from explicit coarse-graining of active particle models that do not have TRS~\cite{Tailleur:08}. In this paper, as is normal in the Model B literature, we shall assume the mobility $M$ to be $\phi$-independent, and set it to unity hereafter. The technical advantages of avoiding multiplicative noise are substantial.
 This simplification is generally considered to be harmless because of the universality of the zero-temperature fixed point, although this can be broken by singular choices of $M[\phi]$~\cite{bray1995lifshitz}. 

Detailed balance is broken in active matter, so that time-reversal symmetry, under which equilibrium Model B is built,  should not longer be imposed on the continuum model at the level of (\ref{eq:MB},\ref{eq:MBJ}). To describe active phase separation, we need to extend Model B to include terms that break TRS. Allowing that the noise remains non-multiplicative, the mobility constant, and that there are no external fields coupled to gradients of $\phi$ (so that rotational symmetry is broken spontaneously if at all), the first nonlinear terms with such a property arise as contributions to $\partial_t\phi$ at order ${\mathcal O}(\nabla^4\phi^2)$. This is because odd powers of $\nabla$ are eliminated by isotropy and terms $\nabla^2\phi^n$ can be absorbed into the passive local free energy $f(\phi)$. Note that any term that can be absorbed by redefining ${\cal F}$ cannot break TRS.

The field theory we study, which will be called Active Model B+ (AMB+), is then obtained from Model B by adding all terms to order $\mathcal{O}(\nabla^4\phi^2)$, while retaining  constant $M = 1$ and $D$. It can be written
\qq
\p_t\phi &=& -\nabla\cdot\left(\mathbf{J}+\sqrt{2D M}\mathbf{\Lambda}\right) \,,  \label{eq:AMB+}\\
\bfJ/M       &=& -\nabla \left[\frac{\delta \mathcal{F}}{\delta\phi} +\lambda|\nabla\phi|^2\right]  + \zeta (\nabla^2\phi)\nabla\phi \,.  \label{eq:AMB+J}
\qqq
Here arise two TRS breaking terms, in $\lambda$ and $\zeta$. Additionally, within $\mathcal{F}$, the coefficient of the square-gradient term in (\ref{eq:F}) is modified to this order as
$K\to K(\phi)=K+2K_1\phi$.
AMB+ is the most general field theory describing the diffusive dynamics of a single scalar field
to order $\mathcal{O}(\nabla^4\phi^2)$, once the choice to keep constant mobility and additive noise has been made. 
(We must of course retain the term arising from the quartic part of the local free energy which is of order $\mathcal{O}(\nabla^2\phi^3)$.)
Any other term at order $\mathcal{O}(\nabla^4\phi^2)$ can be written as a linear combination of $\lambda,\zeta$ and $K_1$ terms.

The way TRS is broken in AMB+ is by destroying the free energy structure that underlies the equilibrium dynamics of Model B without altering the relation between mobility and noise. As shown below (and, for the $\lambda$ term only, in \cite{Wittkowski14,stenhammar2013continuum}) this is what happens when one explicitly coarse-grains interacting active particle models. A quite different route would be to break the fluctuation-dissipation relation between noise and mobility while leaving the free energy structure intact~\cite{paoluzzi2016critical}. This alternative is interesting, but thus far we do not see a similar microphysical motivation for it.

There is a fundamental difference between $K_1$, which merely modifies $\sigma_{eq}$ and does not break detailed balance, and the $\lambda, \zeta$ terms, which locally break TRS. 
The $\lambda$ term can be thought of as locally defining a non-equilibrium chemical potential because, although $\lambda|\nabla\phi|^2$ cannot be written as a derivative of any free energy functional, the resulting current remains of gradient form: ${\bf J} = -\nabla \mu[\phi]$.
The case of $\lambda\neq0$ and $\zeta=0$ is Active Model B (AMB), derived and studied previously~\cite{Wittkowski14,stenhammar2013continuum}. The qualitative physics of AMB is very similar to the one of passive Model B; in particular there is no microphase separation. Instead, the effect of $\lambda\neq 0$ is to change quantitatively 
the coexisting vapor and liquid densities with little or no qualitative change in coarsening behavior~\cite{Wittkowski14}. 

The full AMB+ model (\ref{eq:AMB+},\ref{eq:AMB+J}), and in particular the effect of the $\zeta$ term, remains unstudied until now in the context of phase equilibria and phase ordering kinetics. As we noted in earlier work~\cite{nardini2017entropy}, the $\zeta$ term is distinctive as it allows $\nabla\wedge\bfJ\neq 0$ so that, even in steady state, circulating real-space currents are possible. 
These are prohibited by detailed balance in all passive systems. They are separately prohibited in AMB ($\zeta = 0$), despite broken TRS, by the gradient form of the current in that case. 

At first sight one would not expect circulating current to play a role in phase equilibria or phase ordering because the rotational part of ${\bf J}$ is (by definition) divergence-free, and hence does not enter the equation of motion \eqref{eq:AMB+} for $\phi$.
Importantly though, when $\zeta\neq0$, although via Helmholtz decomposition
one can always define a nonequilibrium chemical potential $\mu$ such that $\nabla\cdot\bfJ=-\nabla^2\mu[\phi]$, the resulting $\mu[\phi]$ becomes {\em nonlocal}. This is because Helmholtz decomposition and the gradient expansion do not commute. Nonlocality of $\mu[\phi]$ will play a central role in the calculations below.

Before presenting those calculations, we note various further points.
$(i)$ Since it respects time reversal symmetry, our $K_1$ term should be relatively unimportant; we focus hereafter on $K_1=0$ but confirm this expectation in Appendix~\ref{K1}.
$(ii)$ Reasons for choosing a symmetric local potential $f(\phi) = f(-\phi)$ in Model B were given above and hold to quartic order, creating an invariance under $(\phi,\lambda,\zeta)\to -(\phi,\lambda,\zeta)$. This duality reduces by half the analytic and numerical calculations required; we exploit it extensively. But it is not fundamental, and we would expect no qualitative changes when using a non-symmetric local free energy $f$. $(iii)$ Although we have 
introduced AMB+ on the basis of conservation laws, symmetry arguments and the gradient expansion, we show in Section \ref{explicit-coarse-graining} that, alongside $\lambda$~\cite{stenhammar2013continuum}, the $\zeta$ term naturally emerges by explicit coarse-graining of particle-based models.
$(iv)$ In~\cite{nardini2017entropy}, a field theory containing all terms up to order $\mathcal{O}(\nabla^4\phi^3)$ was defined (but not studied); our AMB+ model is a special case of it (with, in the notation of that paper, $\kappa_1 = \nu = \lambda_1 = h = 0$). 

In the following, we study AMB+ with analytical tools and simulations in $d=2$; our numerical methods are detailed in Appendix~\ref{simulations}. Unless otherwise specified, simulations are performed with $A=1/4$ and $K=1$. Analytical results are valid in any dimension $d$ unless otherwise specified.

\section{Phase equilibria and the pseudo-tension}\label{binodals}
\begin{figure}
\begin{centering}
\includegraphics[width=1\columnwidth]{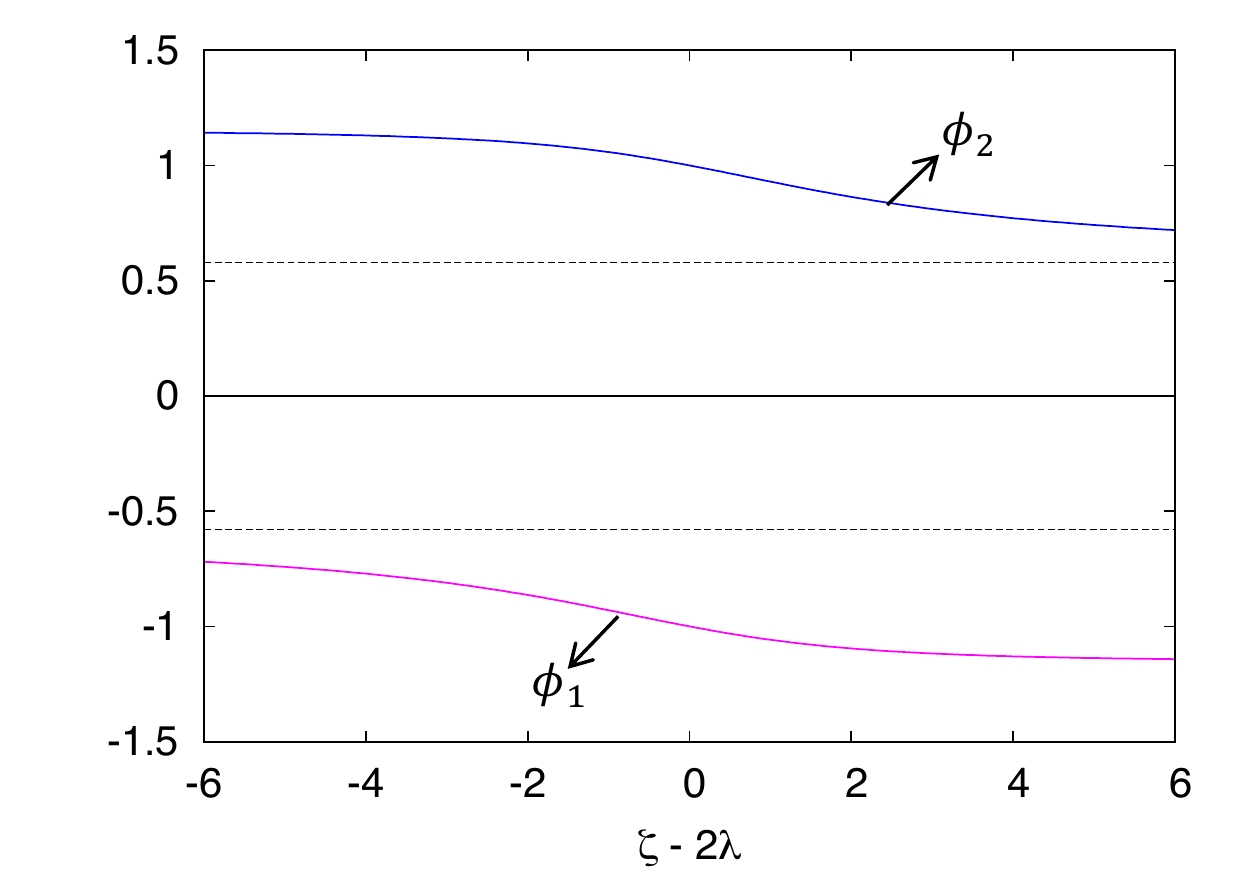}
\par\end{centering}
\caption{The binodals $\phi_1$ and $\phi_2$ (coexisting densities for a flat interface) for $K=1$ and any value of $A$ in the mean-field limit ($D=0$). 
The binodals only depend on one combination of the two activity parameters ($\zeta-2\lambda$) and not on these separately. 
Dashed lines indicate spinodal density defined by $f''(\phi_s)=0.$}
\label{fig:binodals}
\end{figure}

We start by considering the mean-field limit of AMB+ ($D=0$). In this Section, we compute the binodals $\phi_1$ (of the vapor phase) and $\phi_2$ (of the dense phase), defined as the coexisting densities for a flat interface at steady state. Then, we consider phase-separated stationary solutions with spherical symmetry and compute the corrections to the binodals $\phi_{\pm}(R)$ as a function of the droplet radius $R$. Our analysis generalises to AMB+ tools developed for AMB and other models whose currents take gradient form in~\cite{Wittkowski14,solon2018generalized,solon2018generalized2}. It lays the foundations for the main analytic results of this paper, presented in Section \ref{Ostwald-ripening}, where we calculate how $\lambda$ and $\zeta$ affect the Ostwald process.

Consider a flat interface. Clearly (without noise) the density profile depends only on the normal coordinate $x$ and the problem is effectively one-dimensional. It  is then easy to see that the $\zeta$ term and the $\lambda$ term are interchangeable since $\nabla(\nabla\phi)^2 = 2 (\nabla^2\phi)\nabla\phi$.
Hence $\phi_{1,2}$ only depend on the combination of activity parameters $\zeta-2\lambda$. More explicitly, in $d=1$, the deterministic current in (\ref{eq:AMB+J}) is simply $J=-\p_x\mu$, with
\qq\label{eq:stationary-AMB+explicit-flat}
\mu = f'(\phi) -K \phi'' + \left(\lambda-\frac{\zeta}{2}\right)\phi'^2\,
\qqq
where (as previously) $f'(\phi)$ denotes the derivative of $f$ with respect to $\phi$, while $\phi'=\p_x\phi$.

We look for the stationary solution $\phi(x)$ subject to boundary conditions $\phi(-\infty)=\phi_1$ and $\phi(+\infty)=\phi_2$ with a smooth interface around $x=0$. 
Setting $J=0$ implies that the chemical potential $\mu$ equals a constant value $\bar\mu$, fixed by boundary conditions. Then
\qq
\bar\mu = f'(\phi_1) = f'(\phi_2) \,. \label{eq:mu-flat}
\qqq
Equation (\ref{eq:mu-flat}) alone is not enough to determine the binodals $\phi_1$ and $\phi_2$. In the passive limit ($\lambda=\zeta=0$), a second condition is obtained by demanding equal thermodynamic pressure in the two bulk phases: 
 $P_{eq}(\phi_1)=P_{eq}(\phi_2) \label{eq:P-eq}$, where the equilibrium bulk pressure is $P_{eq}(\phi)=\phi f'(\phi) - f(\phi)$.
The same condition can also be derived, without reference to thermodynamics, by multiplying (\ref{eq:mu-flat}) by $\phi'$ and integrating across the interface. 

A formal generalisation of the above result to AMB (and hence AMB+ so long as the interface is flat) was developed in~\cite{solon2018generalized}. This amounts to introducing a pseudo-density $\psi(\phi)$ and a pseudo-potential $g(\phi)$ satisfying:
\qq\label{eq:def-psi-g}
\frac{\p^2 \psi}{\p\phi^2} = \frac{\zeta-2\lambda}{K}\frac{\p \psi}{\p\phi} \quad\text{and}\quad \frac{\p g}{\p \psi} = \frac{\p f}{\p \phi} \,
\qqq
such that $\psi\to\phi$ and $g(\phi) \to f(\phi)$ in the passive limit ($\lambda\to0$ and $\zeta\to0$).
Indeed, with the above definitions, multiplying (\ref{eq:mu-flat}) by $\p_x \psi$ and integrating across the interface gives
\qq
P(\phi_1) = P(\phi_2)   \label{eq:pressure-flat2}
\qqq
where the bulk pseudo-pressure $P(\phi)$ is defined in terms of $\psi(\phi)$ and $g(\phi)$ as 
\qq
P(\phi) = \psi(\phi)\bar\mu - g(\phi) \,. \label{eq:pseudo-pressure}
\qqq

Thus the stationary conditions for finding the binodals $\phi_{1,2}$ is to equate the bulk chemical potentials (\ref{eq:mu-flat}) and 
the pseudo-pressures (\ref{eq:pressure-flat2},\ref{eq:pseudo-pressure}) on both sides of the interface. Explicit results are easily obtained numerically 
because the solutions to  (\ref{eq:def-psi-g}) can be found explicitly. Details are given in Appendix \ref{app:binodals} and the results shown in Fig.~\ref{fig:binodals}. Notice that, as first discovered in~\cite{Wittkowski14} for AMB, only in the passive limit ($\zeta=\lambda=0$) are the binodals at $\phi = \pm1$, which are the minima of the free energy density $f(\phi)$.

Despite its role in phase equilibria, $P$ has no direct link to the mechanical pressure~\cite{Wittkowski14,solon2018generalized}. 
Indeed it appears so far to have no direct physical interpretation; the same applies to our other pseudo-quantities $\psi$, $g$ and (as introduced below) $\sigma$. 
Their role instead is to exactly convert the new mathematics of active phase separation into forms that closely resemble the passive case, allowing
familiar lines of argument  to be applied here (albeit to give unfamiliar outcomes). 

This is a significant achievement since those lines of argument took many years to perfect~\cite{Bray}, and without the transformation to pseudo-variables we could be waiting equally long for answers here. We will see below that our main results indeed match the passive ones under this transformation and also are more easily read in transformed variables than in the original ones.

\begin{figure}
\begin{centering}
\includegraphics[width=1\columnwidth]{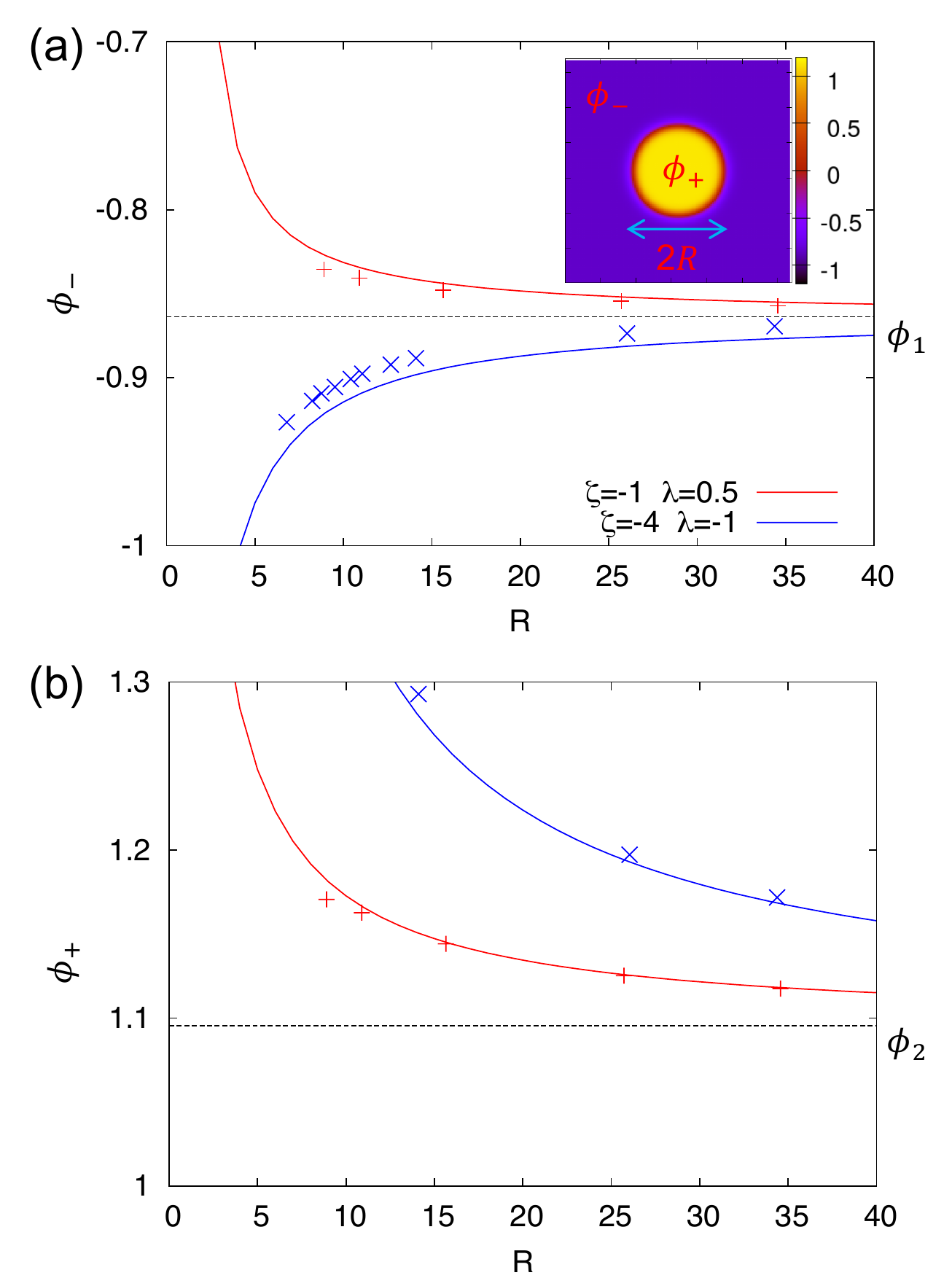}
\par\end{centering}
\caption{ 
Stationary values of the density for a spherical dense droplet in a dilute environment in the mean-field approximation ($D=0$). We plot in (a) the inner density $\phi_+(R)$ and in (b) the outer density $\phi_-(R)$ as a function of droplet radius $R$ for two different sets of activity parameters: $\zeta=-1$ and $\lambda=0.5$ (red, corresponding to region A of Fig.~\ref{fig:phase-diagram-MF}a) and $\zeta=-4$ and $\lambda=-1$ (blue, corresponding to region B of Fig.~\ref{fig:phase-diagram-MF}a).
Points indicate results from the numerical simulations and lines indicate analytical results. When the pseudo surface tension $\sigma$ is positive (red line), $\phi_{\pm}(R)$ are both decreasing functions of $R$. Instead, when $\sigma<0$ (blue line), $\phi_{-}(R)$ is increasing when $R$ increases.}
\label{fig:phi+-}
\end{figure}

We next generalize the same approach to curved interfaces, considering a dense droplet in a dilute environment (see inset of Fig.~\ref{fig:phi+-}(a)). 
The case of a dilute bubble in a dense environment does not need to be analysed separately, as we recall that AMB+ is symmetric under the duality relation $(\phi,\lambda,\zeta)\to -(\phi,\lambda,\zeta)$.

Let us consider a stationary state given by a spherically symmetric droplet of radius $R$ with bulk density $\phi_+(R)$ inside the droplet and $\phi_-(R)$ outside. Although metastable in the large system limit when the noise is switched on ($D\neq 0$) this configuration is, once $\phi_{\pm}$ are chosen correctly, a true stationary state of the mean-field dynamics.
Denoting as $r=|\bfr|$ the radial coordinate from the centre of the droplet,
we are interested in the stationary density profile $\phi(r)$ subject to boundary conditions $\phi(0)=\phi_+$ and $\phi(+\infty)=\phi_-$, where $\phi_+>\phi_-$. 
We expect a smooth interfacial profile around $r=R$ and are mainly interested in situations where $R$ greatly exceeds the interfacial width.

It will be crucial for what follows that, for $d>1$, the $\zeta$ term in the current (\ref{eq:AMB+J}) can be written \emph{via} Helmholtz decomposition as: 
\qq
\mathbf{J}_{\zeta} = \zeta (\nabla^2\phi)\nabla\phi = -\nabla \mu_{\zeta} + \nabla\wedge \mathbf{A} \,.
\qqq
Here the effective chemical potential related to $\bfJ_{\zeta}$ is
\qq
\mu_{\zeta}(\bfr) =  -\int d\bfr' \,(\nabla\cdot \mathbf{J}_{\zeta})(\bfr') \, \nabla^{-2}(|\bfr-\bfr'|)\, \label{eq:mu-zeta}
\qqq
with $\nabla^{-2}(|\bfr-\bfr'|)$ the Green function of the Laplacian. 
Observing that only the gradient part of $\mathbf{J}_{\zeta}$ affects the dynamics of $\phi$, we can forget about ${\bf A}$ in the following. Crucially however, this construction comes at the price of a chemical potential $\mu_{\zeta}[\phi]$ that is non-local.

The full nonequilibrium chemical potential $\mu$, {\em defined} via $ \nabla^2\mu\equiv -\nabla\cdot{\bf J} $, thus 
contains contributions from the passive ($\delta {\mathcal F}/\delta\phi$) and active terms ($\lambda$ and $\zeta$):
\qq
\mu[\phi] = \frac{\delta\mathcal F}{\delta \phi} +\lambda|\nabla\phi|^2 +\mu_{\zeta} \, \label{eq:mu-total}
\qqq
which, using spherical symmetry, can be rewritten as 
\qq
\mu &=& f'(\phi) - K \phi'' - \frac{(d-1)K}{r} \phi' + \Big(\lambda-\frac{\zeta}{2}\Big)\phi'^2  \nonumber \\
       &+& (d-1)\zeta \int_r^\infty \frac{\phi'^2(y)}{y} \,dy \,.  \label{eq:mu-total-curv}
\qqq
Here $\phi'=\p_r\phi$.
At steady state $\mu$ is a constant
 which we shall call $\mu_I(R)$.
Using the boundary conditions $\phi(0)=\phi_+$ and $\phi(+\infty)=\phi_-$, we have
\qq
\mu_I(R) = f'(\phi_-) = f'(\phi_+) + \Delta\,  \label{eq:mu-eq-curv}
\qqq
where
\qq
\Delta &=& (d-1)\zeta \int_0^\infty \frac{\phi'^2(y)}{y} \,dy \label{eq:mu-jump}\\   
          &=& \frac{(d-1) \zeta}{R}\int_0^\infty \phi'^2(y) \,dy
+\mathcal{O}\left(\frac{1}{R^2}\right)\,.   \label{eq:mu-jump2}
\qqq
To go from (\ref{eq:mu-jump}) to (\ref{eq:mu-jump2}) we have assumed a sharp interface limit ($R\gg\xi_{eq}$) so that we can approximate $1/r$ as $1/R$.
Equation (\ref{eq:mu-eq-curv}) states that the full chemical potential $\mu$ inside the droplet is equal to that outside so that 
no current flows across the interface at steady state.
However, the equilibrium part of the bulk chemical potential $\mu_{eq}(\phi)=f'(\phi)$ has a jump $\Delta$ as we cross the interface. This jump is present to cancel one in the $\mu_\zeta$ term whose nonlocal form \eqref{eq:mu-zeta} resembles a Coulomb integral with a curvature-dependent charge density on the interface. Such a charge configuration would create an interfacial step discontinuity in the electrostatic potential and the same happens here for $\mu_\zeta$ in the sharp interface limit. 

As we did for a flat interface, we proceed to find the second condition for $\phi_+$ and $\phi_-$. This is done by  multiplying (\ref{eq:mu-eq-curv}) by $\p_r \psi$, integrating across the interface, and again assuming that $R$ is much larger than the interfacial width. We get
\qq
P(\phi_+) = P(\phi_-) + \frac{(d-1)\sigma}{R} + \mathcal{O}\left(\frac{1}{R^2}\right) \label{eq:press-eq-curv2}
\qqq
where $\sigma$ is a pseudo-tension defined by
\qq
\sigma = \frac{K}{\zeta-2\lambda}\Big[\zeta\mathcal{S}_0 - 2\lambda \mathcal{S}_1\Big] \label{eq:sigma-tilde}
\qqq
defined in terms of two constants that depends on the full shape of the interface:
\qq
\mathcal{S}_0 &=& e^{\frac{\zeta-2\lambda}{K}\phi_2}\,\int_0^\infty \phi'^2(y) \,dy  \label{eq:S0}\\
\mathcal{S}_1 &=& \int_0^\infty  \phi'^2(y) e^{\frac{\zeta-2\lambda}{K}\phi(y)} \,dy\,. \label{eq:S1}
\qqq

The pseudo-pressure balance (\ref{eq:press-eq-curv2}) is formally very similar to what one has in equilibrium, where the Laplace pressure jump crossing the interface is  $\Delta P_{eq}=\frac{(d-1)\sigma_{eq}}{R}$, where $\sigma_{eq}=K\int_0^{\infty} \phi'^2(y) dy$ is the equilibrium interfacial tension. Equation (\ref{eq:press-eq-curv2}) is therefore exactly what one would obtain from the classical equilibrium results upon replacing the pressure and the surface tension with their pseudo counterparts. Moreover, in the passive limit ($\zeta\to0$ and $\lambda\to0$), our results reduce to the equilibrium ones, and in particular $\sigma \to  \sigma_{eq}$. A related analysis of the pseudo pressure applicable to AMB (but not AMB+) was presented  in~\cite{solon2018generalized2}.

The coexisting densities for a spherical droplet of radius $R$ in a dilute environment are given by equations (\ref{eq:mu-eq-curv}) and (\ref{eq:press-eq-curv2}).
However, no explicit values of $\phi_+(R)$ and $\phi_-(R)$ can be found unless $\mathcal{S}_0$ and $\mathcal{S}_1$ are known, which depends on the precise shape of the interface. 
In Appendix~\ref{perturbative}, we explain how to obtain the shape of the interface perturbatively in $1/R$.
Doing so, and imposing the conditions (\ref{eq:mu-eq-curv}) and (\ref{eq:press-eq-curv2}), 
we compute perturbatively the values of $\phi_{\pm}$ up to corrections of order $\mathcal{O}(1/R^2)$. 
These results, which become exact in the sharp interface limit, are shown in Fig. \ref{fig:phi+-} (lines) along with results from mean-field numerical simulation of AMB+. There is good agreement between the two, given the relatively modest values of $R$ simulated.

We are finally able to derive the most important result of this Section. In equilibrium ($\lambda=\zeta=0$), the interfacial tension $\sigma_{eq}$ is always positive. However, in AMB+, $\sigma$ can become negative; from \eqref{eq:sigma-tilde} this happens when 
\qq
|\zeta-2\lambda| <-\zeta \left| 1 - \frac{{\cal S}_0}{{\cal S}_1}  \right|\,.  \label{eq:mean-field3}
\qqq
This is the criterion for $\sigma < 0$ for a dense liquid droplet in a vapor environment. The analogous condition for a vapor bubble
surrounded by the dense liquid then follows from the duality transformation $(\phi,\lambda,\zeta)\to -(\phi,\lambda,\zeta)$.

\begin{figure}
\begin{centering}
\includegraphics[width=1.0\columnwidth]{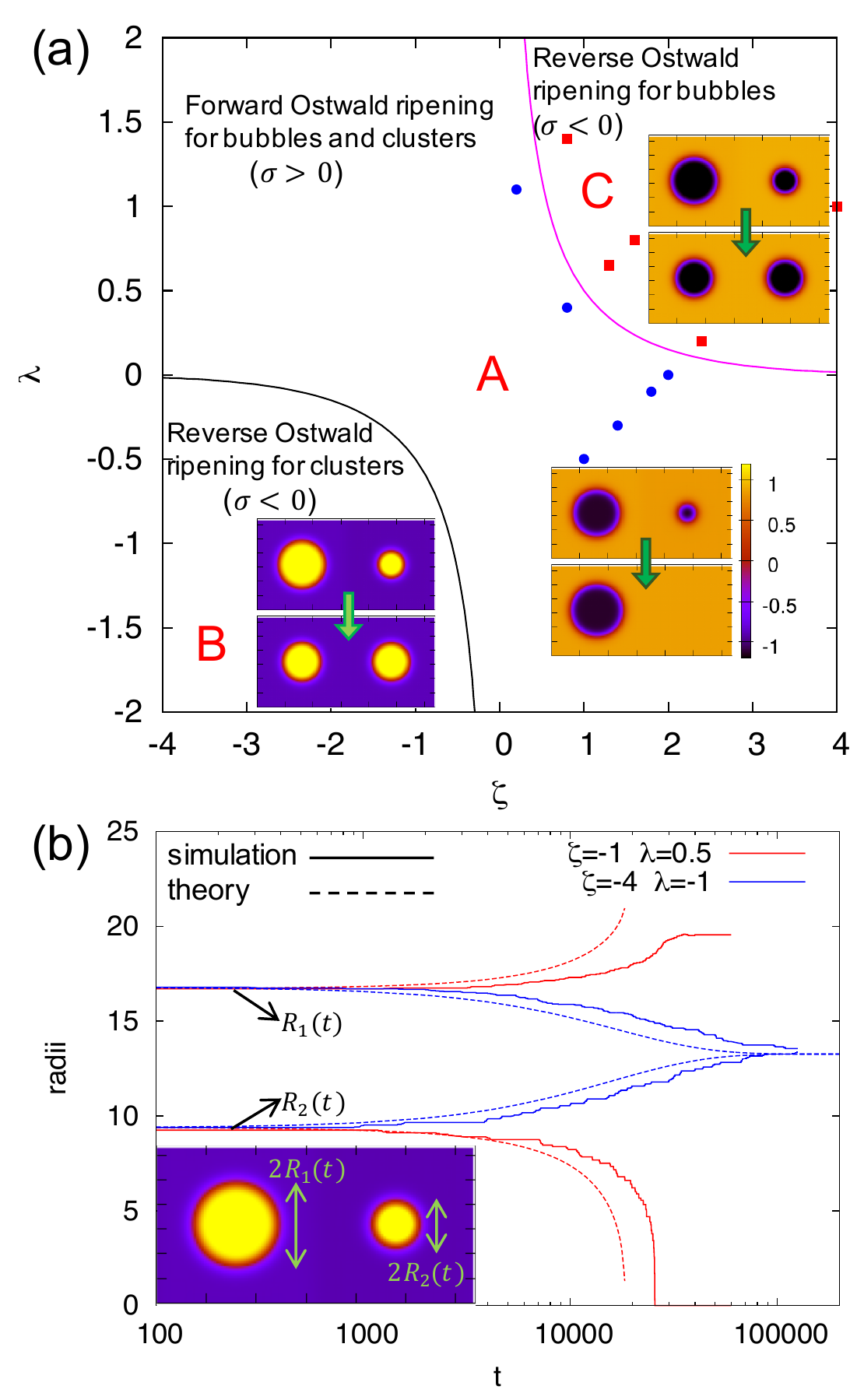}
\par\end{centering}
\caption{ 
(a) Mean field phase diagram of AMB+ with $K=1,K_1=0$ and any value of $A>0$, in $d=2,3$.
Ostwald ripening is reversed where $\sigma<0$.
In region A,  $\sigma>0$, and we have normal/forward Ostwald ripening for both bubbles and droplets.
In region B ({\em resp.}~C) Ostwald ripening is reversed for dense droplets dispersed in a dilute environment ({\em resp.}~dilute bubbles in a dense environment).
Points are results from simulations of two droplets (red/square for reverse and blue/circle for forward Ostwald ripening).
(b) Evolution of two droplets with initial radii $R_1>R_2$ for 
$(\zeta,\lambda)=(-1,0.5)$ (red), corresponding to region A, 
and $(-4,-1)$ (blue), corresponding to region B.
Solid lines are direct numerical simulations of (\ref{eq:AMB+},\ref{eq:AMB+J}) with $D=0$ and dashed lines are predictions from our theory (see Appendix \ref{app:two-droplets}).
We observe forward Ostwald ripening in the first case and reverse Ostwald ripening for the second, as predicted.
\label{fig:phase-diagram-MF}}
\end{figure}

The phase diagram is thereby divided in three regions, as represented in Fig.~\ref{fig:phase-diagram-MF}(a), as follows. In region A, $\sigma>0$ regardless of the sign of the interfacial curvature. Note that AMB ($\zeta = 0$) lies entirely within this region. In region B, $\sigma<0$ for liquid droplets in a vapor environment but $\sigma>0$ for vapor bubbles in a liquid environment. In region C, conversely, $\sigma$ is negative for vapor bubbles and positive for liquid droplets.

We have shown that the pseudo-tension $\sigma$ determines the pseudo-pressure jump at a curved interface in the same way that, in equilibrium systems, the usual tension $\sigma_{eq}$ determines the Laplace pressure jump. However, this does not mean that $\sigma$ has any direct connection to the mechanical tension as defined for active systems \emph{e.g.}~in~\cite{bialke2015negative}. Crucially, the presence of a negative pseudo-tension does not cause interfaces to become unstable: the coexisting densities $\phi_{\pm}(R)$ remain well defined.  The effect of negative $\sigma$ is instead to change how these quantities depend on $R$. 
 In equilibrium systems, both $\phi_{\pm}$ are decreasing functions of $R$. This remains true in AMB+ for $\sigma>0$, as shown for a particular choice of $\lambda$ and $\zeta$ within region A, in Fig.~\ref{fig:phi+-}. 
 Instead, when $\lambda$ and $\zeta$ belong to region B, while $\phi_+(R)$ is still a decreasing function of $R$, $\phi_-(R)$ now increases with $R$. 

\section{Reverse Ostwald ripening: Mean-field phase diagram} \label{Ostwald-ripening}
In this Section we show how negative pseudo-tension affects the Ostwald process, which is the kinetic pathway to bulk liquid-vapor phase separation in equilibrium as described by Model B. As in the previous Section, we focus on region B of Fig.~\ref{fig:phase-diagram-MF}(a); the behavior in region 
C then follows by the duality relation.

A dramatic change in behavior on passing from region A to region B in Fig.~\ref{fig:phase-diagram-MF}(a) is natural because, as explained in Section \ref{binodals} and in Fig.~\ref{fig:phi+-}, this changes the sign of $\p_R \phi_-(R)$. 
In region A, just as in equilibrium models, $\sigma>0$ giving normal or forward Ostwald ripening in which the density outside a small droplet is higher than that outside a bigger droplet and thus the current flows from small droplets to big ones. In contrast, in region B,  $\sigma<0$ and the density outside a small droplet is lower than that outside a bigger droplet; the particle current now flows from big to small droplets. This is the regime of reverse Ostwald ripening in which small droplets grow at the expense of larger droplets, which shrink.

We now make the above reasoning precise by generalizing to AMB+ the standard arguments for Ostwald ripening in equilibrium models~\cite{Bray,CatesJFM:2018}. 
We work in $d=3$ dimensions; for $d=2$ there are correction terms to the following arguments involving the logarithm of the total phase volume occupied by droplets, but these do not change our conclusions (see, \emph{e.g.,}~\cite{alikakos2004ostwald}).  

We consider a spherical droplet of radius $R$ and denote as $r=|\mathbf{r}|$ the distance from the centre of this droplet. On a rapid time scale the densities just inside and outside the droplet relax to their quasi-stationary values, $\phi_+$ and $\phi_-$, which differ from the binodal densities $\phi_2$ and $\phi_1$ respectively by terms of order $1/R$.
Now we assume the droplet to be surrounded by a sea of distant droplets, such that the mean density of the vapor phase at $r=\infty$ is $\phi_s=\phi_1+\epsilon(t)$, 
where $\epsilon$ is the supersaturation.
We wish to calculate the time evolution of the radius $R(t)$ of our central droplet; this depends on $\epsilon(t)$.
As is standard \cite{Bray} our argument assumes a quasi-static diffusion so that $\nabla^2\mu(r)=0$ at any time. 
This is to be solved with boundary conditions $\mu(\infty)=f'(\phi_s)$ and $\mu(R)=\mu_I(R)$.
Once the resulting solution $\mu(r)$ is known, mass conservation allows us to give an equation for the temporal evolution of the droplet radius $R(t)$.
 
Rewriting (\ref{eq:press-eq-curv2}) with $\phi_{\pm}=\phi_{2,1}+\mathcal{O}(1/R)$, we find that the chemical potential at the interface of the droplet is:
\qq
\mu_I(R) = \frac{\Delta g}{\Delta \psi}  + \frac{(d-1)\sigma}{\Delta \psi \,R} + \mathcal{O}\left(\frac{1}{R^2}\right) \,,  \label{eq:muR}
\qqq
where $\Delta g=g(\phi_2)-g(\phi_1)$, $\Delta \psi = \psi(\phi_2)-\psi(\phi_1)$,  and $\sigma$ is the pseudo-tension defined in (\ref{eq:sigma-tilde}).
Note that for $\lambda=\zeta=0$, one has $g=f$, $\psi = \phi$ and $\sigma = \sigma_{eq}$, thereby recovering the classical result for equilibrium models~\cite{Bray}.

Next we solve $\nabla^2\mu=0$ subject to boundary conditions $\mu(r=R)=\mu_I(R)$ and $\mu(r=\infty)=f'(\phi_s)$ to obtain the quasi-static nonequilibrium chemical potential:
\qq
\mu(r) = \frac{R(\mu_I(R)-f'(\phi_s))}{r} + f'(\phi_s)\,.
\qqq 
The inward radial current at the interface of the droplet $J_I$ now obeys $-J_I=-\nabla\mu\big|_R=(\mu_I(R)-f'(\phi_s)){\bf\hat{r}}/R$.
Finally using the conservation of mass:
\qq
\frac{d}{dt}\left( \frac{4}{3}\pi R^3(\phi_+-\phi_-) \right) = J_I 4\pi R^2 \, ,
\qqq
we obtain the time evolution of the droplet radius $R(t)$:
\qq\label{eq:R-time-evolution}
\dot{R} = 
 \frac{\beta}{R\Delta\phi}  
\left[ \frac{1}{R_s} -\frac{1}{R} \right] +\mathcal{O}\left(\frac{1}{R^3}\right)
\qqq
where $R_s$ is a fixed-point radius, $\beta$ is a rate parameter, and $\Delta\phi = \phi_2-\phi_1$.
The explicit expressions for $R_s$ is:
\qq
R_s = \frac{\Delta \psi \,\beta}{f'(\phi_s)\Delta \psi - \Delta g}  \label{eq:Rs}
\qqq
and likewise for $\beta$:
\qq
\beta = \frac{(d-1){\sigma}}{ \Delta \psi} \,.  \label{eq:beta}
\qqq

In equilibrium models, the surface tension $\sigma\to \sigma_{eq}$ is always positive and thus $\beta$ and $R_s$ are also positive.
$R_s$ is then an unstable fixed-point radius:
droplets smaller than $R_s$ shrink and droplets larger than $R_s$ grow.
This conventional Ostwald process is maintained, for droplets and bubbles alike, throughout region A of the phase diagram of AMB+, as shown in Fig.~\ref{fig:phase-diagram-MF}(a), where $\sigma>0$.

In AMB+, however, the pseudo-tension $\sigma$ is negative when condition (\ref{eq:mean-field3}) is met, which implies that the rate $\beta$ is also negative.
To determine the sign of $R_s$, we Taylor expand Eq.~(\ref{eq:Rs}) for small $\epsilon$ to obtain:
\qq
\frac{\beta}{R_s} = \frac{f''(\phi_1)}{\Delta\phi\Delta \psi}\epsilon +\mathcal{O}(\epsilon^2)
\qqq
where we have used the fact that $\phi_{1,2}$ satisfy (\ref{eq:mu-flat}) and (\ref{eq:pressure-flat2}). 
Since the binodal $\phi_1$ is always outside the spinodal, $f''(\phi_1)>0$ is always positive. 
We also note that when $\sigma$ (hence $\beta$) is negative,
the supersaturation $\epsilon$, which is set by the average of $\phi_{-}-\phi_1$ across the population of distant droplets, is necessarily also negative;
see Appendix \ref{app:two-droplets} below, and also compare the red and blue curve in Fig.~\ref{fig:phi+-}(a)). 
We thus conclude that, for small $\epsilon$ in regime B ($\sigma<0$), $R_s>0$ and $\beta<0$. 
$R_s$ is now a stable fixed point radius: droplets of smaller than $R_s$ grow and those larger than $R_s$ shrink.
This is the regime of reverse Ostwald ripening for liquid droplets dispersed in a continuous vapor phase.

Regime C in the mean field phase diagram (Fig.~\ref{fig:phase-diagram-MF}(a))
gives another region where Ostwald ripening is reversed, but now for vapor bubbles dispersed in a dense liquid environment.
This can be derived explicitly as above by considering a bubble of radius $R$ with density $\phi_1$ inside and $\phi_2-\epsilon$ at infinity.
Alternatively, it follows directly from the duality relation of AMB+. Note that there is no choice of parameters for which {\em both} droplets and bubbles are subject to the reverse Ostwald mechanism. In consequence, as confirmed numerically below, we expect microphase separation to occur in AMB+ on one side of the phase diagram or the other but not both.

To test our analytic mean-field predictions, we performed numerical simulations in the mean-field limit ($D=0$) for several values of $\lambda$ and $\zeta$, using as initial condition two droplets with different initial radii $R_1(0)>R_2(0)$.
Fig.~\ref{fig:phase-diagram-MF}(b) shows the typical time evolution of the radii $R_1(t)$ and $R_2(t)$ (solid lines) for parameter values in regions A and B in Fig.~\ref{fig:phase-diagram-MF}(a). In region A we confirmed that the larger droplet grows at the expense of the smaller one whereas in region B the big droplet shrinks until $R_{1,2}$ reach the same value.
For all the other  values of $\lambda$ and $\zeta$ that we also tried, our numerical results for two droplets likewise confirmed the analytic predictions for whether the normal or the reverse Ostwald process should be seen (see points in Fig.~\ref{fig:phase-diagram-MF}(a)).

As shown in Appendix~\ref{app:two-droplets}, the above calculations can be extended to track explicitly the evolution in time of the two radii $R_{1,2}(t)$ in a binary droplet system. (We have not seen a calculation of this kind in the literature even for equilibrium models. To achieve this one needs to assume that the second droplet acts as a distant boundary condition on the first (and vice versa), while retaining radial symmetry. This effective-medium approximation was used above (as is standard \cite{Bray}) for the multi-droplet case, but it is clearly less valid when only two droplets are present.  An additional source of error is that the calculation assumes the two droplets are separated by a distance large compared to their radii which is not really met in simulations due to computational limitations. Given these shortcomings, one does not expect quantitative agreement, but as shown in Fig.~\ref{fig:phase-diagram-MF}(b), these predictions are qualitatively in accord with our two-droplet simulations.

\section{Simulations with finite noise} \label{noise}

\begin{figure*}
\begin{centering}
\includegraphics[width=1.\textwidth]{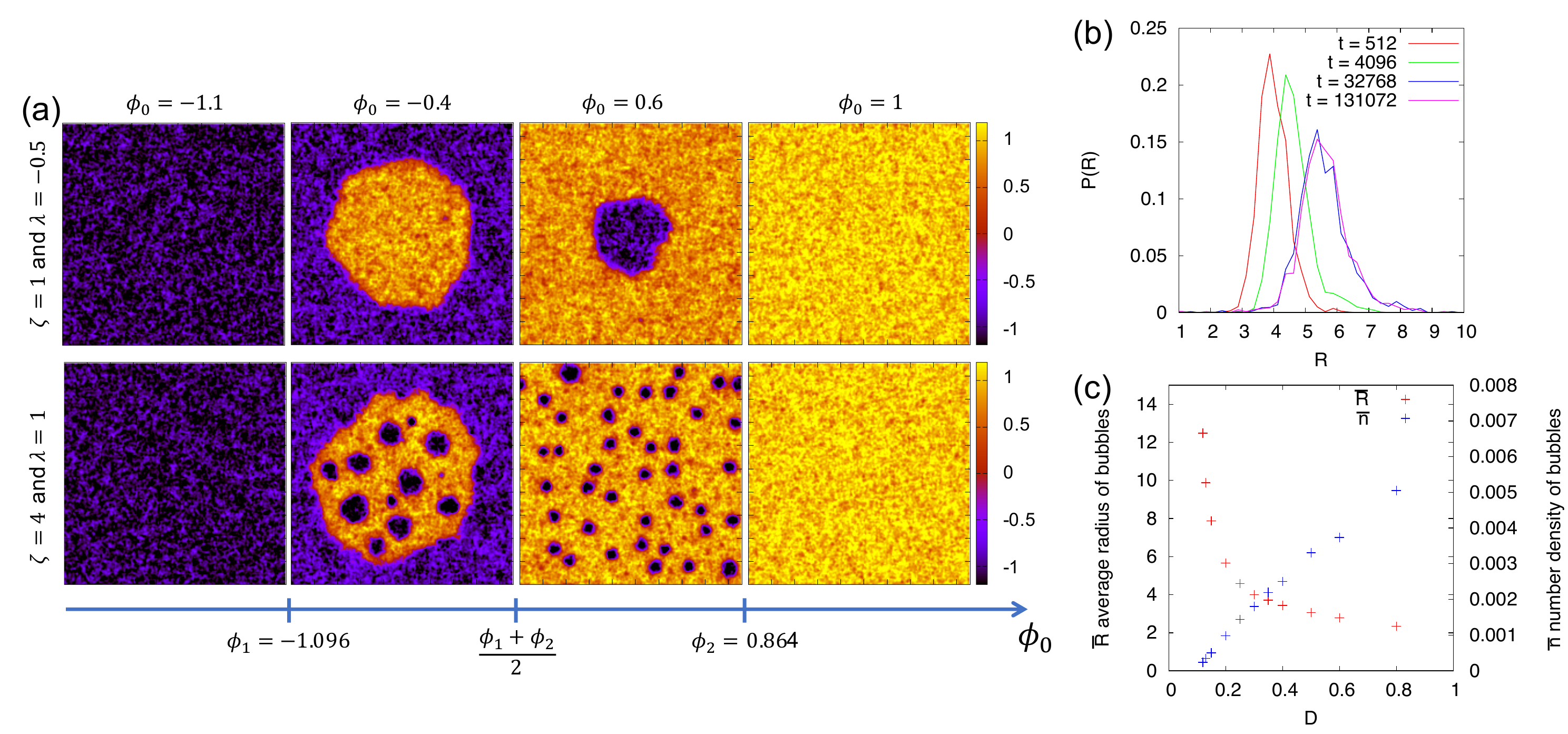}
\par\end{centering}
\caption{ 
(a) Steady state phase diagram as a function of global density $\phi_0=\frac{1}{V}\int d\bfr\phi$ for two different sets of activity parameters:
$\zeta=1,\lambda=-0.5$ corresponding to region A in Fig.~\ref{fig:phase-diagram-MF}(a), and $\zeta=4,\lambda=1$ corresponding to region C. 
In A, the steady state phase diagram resembles equilibrium Model B with full phase separation into dense (yellow) and dilute (blue) phases. 
In C, besides the low and high density uniform phases, we observe either bubbly phase separation or a homogeneous bubble phase. 
Since AMB+ is symmetric under $(\phi,\lambda,\zeta)\to -(\phi,\lambda,\zeta)$, the phase diagram in region B is obtained by exchanging dense with dilute regions.
(b) Probability distribution of bubble radius for parameters $\zeta=4,\lambda=1$ as a function of time,
starting from an initial uniform density $\phi(\bfr,t=0)=\phi_0 = 0.6$.
The system reaches a steady state at $t\simeq32000$.
(c) Average radius of bubbles $\bar{R}$ and number density of bubbles $\bar{n}$ in the homogenous bubble phase as a function of $D$.
Parameters: $\zeta=4,\lambda=1$ and $\phi_0=0.6$.} 
\label{fig:phase-diagram-full}
\end{figure*}

Our mean-field analysis show that the reverse Ostwald process can arise generically in active phase separation, without the need for specific interaction mechanisms or fine-tuning of parameters. It is then natural to speculate that this leads to microphase-separated stationary states, at least when dynamical noise is present so that the steady-state is independent of initial conditions.  Indeed, when noise is present, each time a droplet/bubble is nucleated, it will converge to roughly the same size as the others.
To test this hypothesis, we have performed full stochastic simulations of AMB+ with finite noise $D=0.2$ on a $192\times192$ lattice 
(see Appendix~\ref{simulations} for numerical details).
We consider two sets of activity parameters: $(\zeta=1,\lambda=-0.5)$ and $(\zeta=4,\lambda=1)$ corresponding to regions A and C of the mean-field phase diagram.
These parameters have the same $\zeta-2\lambda$ and thus the same binodals: $\phi_2\simeq0.86$ and $\phi_1\simeq-1.10$.

Fig.~\ref{fig:phase-diagram-full}(a) shows the steady state phase diagram as a function of global mean density $\phi_0=\frac{1}{V}\int d\bfr \, \phi(\bfr,t)$. 
Outside the binodals, $\phi_0<\phi_1$ or $\phi_0>\phi_2$, the uniform phase is always stable.
In region A (top row), the phase diagram resembles that of passive Model B, with bulk phase separation into dense liquid and dilute vapor phases, 
except that the binodals $\phi_{1,2}$ are now shifted. This is the same phenomenology as described previously for AMB~\cite{Wittkowski14}, whose parameters ($\lambda\neq 0,\zeta = 0$) indeed lie within region A.

The phenomenology in region C (bottom row) -- or equivalently by duality in region B -- is much more interesting. Besides the low-density and high-density uniform phases when $\phi_0$ is outside the binodals, for $\frac{\phi_1+\phi_2}{2}\lesssim\phi_0\lesssim\phi_2$ in region C we observe initial coarsening of dilute bubbles. At long times, however, the coarsening arrests and the number of bubbles fluctuates around some average (Supp.~Movie 1).
This represents nonequilibrium microphase separation in the form of a homogeneous bubble phase, which becomes a cluster phase (Fig.~\ref{fig:snapshots}(b)) when we flip into region B by the duality transformation $(\zeta,\lambda,\phi_0)\rightarrow-(\zeta,\lambda,\phi_0)$. We have checked that this homogeneous microphase separation is the true steady steady within the above range of $\phi_0$, by establishing that it is reached not only from a uniform initial state but also one at bulk phase coexistence (Supp.~Movie 2). 

At global densities $\phi_0$ in the range $\phi_1\lesssim\phi_0\lesssim\frac{\phi_1+\phi_2}{2}$ within region C, the system creates a globally inhomogeneous steady state resembling conventional bulk phase separation, except that within the dense phase, vapor bubbles are continuously formed and then expelled to the exterior bulk vapor (Supp.~Movie 3 and 4). This `bubbly phase separation' looks very similar dynamically to what was reported in simulations of active Brownian particles with hard-core repulsions~\cite{stenhammar2014phase}. Given what we have learned about the global phase diagram, it can best be viewed as a bulk coexistence between a homogeneous bubble phase and a conventional vapor. By duality we find a range of densities within region B in which there is bulk coexistence between a cluster phase and dense liquid. We have seen no reports of this in either experiments or particle-based simulations on active particles; but since neither has explored parameter space fully, we predict that such states will be found in future.

The dynamical properties of the homogeneous bubbles phase (and its cluster-phase dual) are intriguing. 
In steady state, bubbles are created by nucleation and destroyed mainly by coalescence with others at low noise (Supp.~Movie 1). They can, however, 
also disappear by shrinking especially at high noise (Supp.~Movie 5). 
The steady state average bubble size $\bar{R}$ is then expected to depend strongly on noise, set by a balance between the noise-induced nucleation, the mean-field dynamics inducing reversed Ostwald process, the diffusion (which leads to coalescence), and the possible shrinking at high noise. Interestingly, see Fig.~\ref{fig:phase-diagram-full}(c), we found that $\bar{R}$ increases when decreasing the noise strength $D$, which is a non-trivial statistical property of this phase. 

We also measured the probability distribution of the bubble radius $P(R)$, see  Fig.~\ref{fig:phase-diagram-full}(b), as a function of time when starting from a uniform initial state. We found that $P(R)$ converges to a well defined shape at long times.
We leave for the future any attempt to predict $\bar{R}$ and $P(R)$ in the uniform bubble/cluster phase. In particular, it would be interesting to generalize the Lifshitz-Slyozov theory~\cite{Bray}, using the mapping to pseudo-pressure and pseudo-tension, to predict $P(R)$. However, the fact that $\bar{R}$ depends on the noise level suggests that this theory, which is strictly mean-field in character, will not suffice when $\sigma<0$.
Since the noiseless two-droplet problem then has a stable solution at equal droplet size, one possibility is that, for any given initial conditions, a noiseless theory at Lifshitz-Slyozov level lands the system somewhere on a manifold of static solutions with strictly monodisperse droplets. So long as they all have the same size, then at effective medium level, such droplets can be in stable equilibrium with each other and may have no reason to evolve further. If so, the selection of a mean droplet size would have to depend on a perturbation to this mean-field scenario, such as dynamical noise. $\bar{R}$  could then be set by a balance between noise and deterministic terms, as speculated in the previous paragraph.

\section{Explicit coarse graining from microscopic dynamics}\label{explicit-coarse-graining}
In Section \ref{model}, we introduced AMB+ on the basis of top-down arguments based on conservation laws and a gradient expansion. Here we show how the form of the TRS-breaking terms in the force density $\bfJ/M$ in \eqref{eq:AMB+J}, crucial for the phenomenology found above, emerge from explicit coarse graining of a microscopic dynamics. More precisely, we obtain by coarse-graining a field theory closely related to AMB+, the main differences being that the mobility $M$ is density-dependent, and the local free energy is not a polynomial. We do not consider these differences important, because they also arise for Model B, where constant $M$ and the polynomial form of $f(\phi)$ are not fundamental properties inherited from some microscopic model, but strategic choices made to simplify the field theory into one of $\phi^4$ form with additive noise only. AMB+ makes the same choices, so our task here is only to check that the $\zeta$ and $\lambda$ terms can emerge as expected from microscopic models.

We start from a model of active Ornstein-Uhlenbeck particles (AOUPs)~\cite{Maggi:14b,Brader:15,fodor2016far} 
where each overdamped particle $i$ experiences 
first an active force $v_0\bfu_i$, where ${\bfu}_i$ is exponentially correlated unit noise with persistence time $\tau$, and 
second a two-body interaction with other particles.
In a similar spirit to liquid-state theory \cite{hansen1990theory}, we treat separately the hard-core part of the pair interaction from its softer tail. 
Following~\cite{Tailleur:08,stenhammar2013continuum,Cates:15} we argue that the hard-core interactions effectively coarse-grain into an effective propulsion speed $v(\bfr,[\rho])$ that depends functionally on the coarse grained density $\rho$ of active particles. 

Differently to previous approaches~\cite{Tailleur08,solon2018generalized}, we keep explicitly the soft pair interaction via a residual two body potential $U= U(|\bfr_i-\bfr_j|)$. 
Thus, with unit damping, the particles' positions $\bfr_i$ and directions $\bfu_i$ obey:
\qq
\dot{\bfr}_i &=& v \bfu_i -\sum_j\nabla_i  U + \sqrt{2D_t}\, \pmb{\xi}_i \, \label{eq:AOUPs-main-1} \\
\dot{\bfu}_i &=& -\bfu_i/\tau + \sqrt{2/\tau}\, \pmb{\eta}_i \, \label{eq:AOUPs-main-2}
\qqq
where $\pmb{\eta}_i$ and $\pmb{\xi}_i$ are independent Gaussian random variables with zero mean and unit variance.
$D_t$ is a passive translational diffusion constant.
An objection to this treatment of  AOUPs is that replacing the hard-core interactions by $v[\rho]$ sacrifices the equation of state between pressure and density that was present for the original model~\cite{solonPhD,Grosberg-PRE2017}. However, this refers to mechanical pressure which plays no part in the analyses of this paper. 

In any case,  
(\ref{eq:AOUPs-main-1}-\ref{eq:AOUPs-main-2}) can equally well be viewed as a microscopic dynamical model in its own right. At this level it describes motile micro-organisms whose exponentially autocorrelated (AOUP-like) self-propulsion has a root-mean-square swim speed $v({\bf r},[\rho])$ that depends on local density via a chemical (quorum-sensing) interaction rather than a mechanical one~\cite{solon2018generalized}. There is in addition a soft pair interaction $U$ between the motile particles. 

We sketch here the coarse-graining procedure, focusing on the effect of $U$, and refer to Appendix~\ref{app:cg} for more details.
Technically, the novelty of our approach stands in first taking the small $\tau$ limit, and then writing an equation of motion for the empirical density $\rho_d(\bfr,t)=\sum_i \delta(\bfr - \bfr_i)$ directly, instead of first writing one for $\psi_d(\bfr,\bfu,t)=\sum_i \delta(\bfr - \bfr_i)\delta(\bfu - \bfu_i)$ and then passing to the hydrodynamic limit to obtain the evolution of $\rho_d$ as is more often done~\cite{solon2018generalized}. Our method is much more compact and gives us the possibility of treating $U\neq 0$, which would pose difficulties via the standard route. Still, for $U=0$, we get exactly the same results as already known in the literature~\cite{Tailleur:08,solon2018generalized}.

It is well known from the theory of slow-fast dynamical systems~\cite{pavliotis2008multiscale,Lau:07} that, in the limit of small $\tau$, $\bfu_i$ can be eliminated from (\ref{eq:AOUPs-main-1},\ref{eq:AOUPs-main-2}) leading to
\qq\label{app:eq:AOUPs-reduced}
\dot{\bfr}_i =_s \sqrt{2\tau}\,v \pmb{\eta}_i  -\sum_j \nabla_i  U + \sqrt{2T}\pmb{\xi}_i\,.
\qqq
where, importantly, the equatily has to be interpreted in the Stratonovich sense, which is the meaning of $=_s$. 
This equation can now be transformed to It\^{o} form, giving
\qq\label{app:eq:AOUPs-reduced-Ito}
\dot{\bfr}_i = v \sqrt{2 \tau}\pmb{\eta}_i -  \tau v \nabla_i  v 
-\sum_j \nabla_i  U + \sqrt{2T}\pmb{\xi}_i\,.
\qqq
We then write the Dean equation~\cite{Dean}, which is the equation formally solved by $\rho_d$ and, finally, we replace $\rho_d$ by its smooth limit $\rho$. Note that only on this replacement is information lost and the coarse graining really done~\cite{nardiniperturbative2016,archer2004dynamical}. Such a procedure can be rigorously shown to give correctly the statistics of $\rho_d$ at least in the case where the two body interaction $U$ is weak and the density large (the so-called Kac limit) \cite{dawsont1987large,nardiniperturbative2016}. More generally, passing from $\rho_d$ to $\rho$ remains a standard procedure in the theoretical physics literature even far from this limit~\cite{kawasaki1994stochastic,barre2015motility,donev2014dynamic}, although its proper justification is then an outstanding mathematical question~\cite{dawsont1987large}.

The resulting dynamics is written as (Appendix~\ref{app:cg}):
\qq\label{eq:AMB+general}
\p_t \rho = -\nabla \cdot\{M[\rho]\tilde\bfJ_{cg} +\sqrt{2M[\rho]} {\bf \Lambda} \}\,,
\qqq 
where 
\qq
M[\rho] &=& \rho (\tau v^2[\rho] + T) \,. \label{eq:app:Dean-Many-AMB+M}
\qqq
The effective force density $\tilde\bfJ_{cg}$ sets the deterministic current $\bfJ_{cg}$ via $\bfJ_{cg}=M[\rho]\tilde\bfJ_{cg} $. The force density due to quorum sensing and to the two-body potential add linearly
\qq
\tilde{\bf J}_{cg}&=& \tilde{\bfJ}_{qs}+ \tilde{\bfJ}_{r} \label{Jcg}
\qqq
where
\qq\label{eq:Dean-Many-AMB+}
\tilde{\bfJ}_{qs}&=&- \nabla \mu_{qs}\\
\mu_{qs} &=& \log \rho +\frac{1}{2}\log \left(\tau v[\rho]^2+T\right)\label{eq:Dean-Many-AMB+muqs}\\
 \tilde{\bfJ}_{r} &=&
  -\frac{\rho \nabla (U\star \rho)}{M[\rho]}\,.\label{eq:Dean-Many-AMB+Jtilde} 
\qqq
Here $\star $ denotes the convolution product and we have explicitly allowed that $v$ is a functional of $\rho$. The subscript $qs$ stands for quorum sensing (see above). 

It is now clear that, if $U=0$, the effective force density $\tilde\bfJ_{cg}$ becomes the gradient of a nonequilibrium chemical potential $\tilde\bfJ_{cg} = \tilde{\bf J}_{qs}= -\nabla \mu_{qs} $, irrespective of the functional dependence of $v$ on $\rho$ and of
whether an equilibrium structure is present (that is, whether $\mu_{qs}=\delta \mathcal{ F}_{qs}/\delta \rho$ for some functional $\mathcal{F}_{qs}[\rho]$).
This special case has been widely considered in the literature \cite{Tailleur:08,Wittkowski14,solon2018generalized}. 
Explicitly, expanding $v[\rho]$ at lowest order in gradients and retaining only terms allowed by rotational symmetry, we write 
$ v[\rho]=v(\rho)+\beta(\rho) |\nabla \rho|^2 +\gamma(\rho) \nabla^2\rho$.
Note that $\beta\neq0$ iff the active particles can detect local gradients directly; the $\gamma$ term instead describes particles whose speed depends on the average density in a small neighborhood \cite{Cates:15}. Expanding (\ref{eq:Dean-Many-AMB+muqs}), we obtain
\qq\label{app:eq:cg-qs-complete}
\mu_{qs}=
\frac{\delta \mathcal{F}_{qs}}{\delta \rho} +\lambda_{qs}|\nabla\rho|^2 +o(\nabla^2\rho^2)
\qqq
where
\qq\label{eq:Fqs-prx}
 \mathcal{F}_{qs} = 
 \int d\bfr\, \left\{f_{qs}(\rho) +
 \frac{K^{qs}+2K_1^{qs}\rho}{2} |\nabla\rho|^2\right\}\,.
\qqq
The expressions of $f_{qs}, K^{qs}, K_1^{qs}$ and $\lambda_{qs}$ in terms of microscopic parameters are given in Appendix \ref{app:cg} and agree with those previously obtained in the literature.

The above coarse-graining at $U = 0$ represents a pure quorum-sensing model. Importantly, this gives only one of the two TRS-breaking terms in \eqref{eq:AMB+J}, namely $\lambda$. It does not give us the $\zeta$ term that distinguishes AMB+ from AMB and is responsible for all the new physics discussed in this paper. But we can now see immediately how the $\zeta$ term appears in $\tilde{\bf J}_{cg}$ in \eqref{Jcg} as soon as the soft potential $U$ is nonzero: it does so because $\tilde{\bfJ}_r$ is, according to \eqref{eq:Dean-Many-AMB+Jtilde}, not the gradient of any scalar function. More explicitly, performing a gradient expansion, we obtain 
\qq\label{eq:cg-Jtilde}
\tilde{\bfJ}_r
=-\nabla \mu_{r} +\zeta_{cg} (\nabla\rho)(\nabla^2\rho)+o(\nabla^3\rho^2)
\qqq
where the explicitly derived $\zeta$ parameter is
\qq\label{eq:zeta-explicit}
\zeta_{cg}= \frac{2\tau v_0 }{(T+\tau v_0^2)^2}(\gamma_0 C - v_1 B)\,.
\qqq
Here $v_0,v_1$ and $\gamma_0$ are defined by $v(\rho)=v_0+v_1\rho+o(\rho)$ and $\gamma(\rho)=\gamma_0+\mathcal{O}(\rho)$. Finally, the constants $B$ and $C$ only depend on the two body potential $U$
\qq\label{app:eq-BC-cg}
  B&=&  -\frac{\Omega}{2d(d+2)}  \int_0^{\infty} dy\,y^{d+2} \,U'(y)\label{app:eq-BC-cg-2}\\
  C&=&-\frac{\Omega}{d}\int_0^{\infty} dy \,y^{d}\,U'(y)\label{app:eq-BC-cg-1}
  \qqq
where $\Omega $ is the surface of the unit sphere. The effective chemical potential $\mu_r$ in (\ref{eq:cg-Jtilde}) has the same form as (\ref{app:eq:cg-qs-complete}) with the subscript $qs$ replaced by $r$ and the explicit dependence of  $f_{r}, K^{r}, K_1^{r}$ and $\lambda_{r}$ given in Appendix \ref{app:cg}.

These results from coarse graining now pass to those of AMB+ via $(i)$ the linear transformation from $\rho$ to $\phi$ \cite{stenhammar2013continuum}; $(ii)$ replacement of the local part of $f_{cg}(\rho)$ by an even quartic polynomial in $\phi$; and $(iii)$ suppression of the density dependence in the mobility $M$, allowing this to be set to unity so that $\bfJ_{cg} = \tilde\bfJ_{cg}$ and the noise becomes additive. As already stressed at the beginning of this Section, these three steps exactly mirror those used when passing from coarse-grained microscopic models for diffusive phase separation in passive systems to their canonical incarnation as passive Model B \cite{Bray}. 

In summary, our explicit coarse graining shows that both of the TRS-breaking terms considered in AMB+ (\ref{eq:AMB+},\ref{eq:AMB+J}) emerge naturally from at least one well-motivated model for the microscopic dynamics of interacting active particles. A more careful study of that coarse-grained model in its own right (without the three additional simplifications listed above) is postponed to future works.

\section{Conclusion} \label{conclusion}
We have presented results for Active Model B+ (AMB+), a field theory that, at leading order in a gradient expansion, fully generalises the canonical model of passive phase separation (Model B) to break time-reversal symmetry. The model has two separate activity parameters, $\lambda$ and $\zeta$. When both are small, Ostwald ripening operates normally and bulk phase separation is retained. It is also retained at large activity when $\lambda$ and $\zeta$ have opposite signs. However, when both are large and positive, one finds a regime of reverse Ostwald ripening for vapor bubbles in a dense liquid; for both large and negative, one finds this for dense droplets in a vapor.  

This unexpected reversal of the Ostwald process is captured mathematically by a negative pseudo-tension which plays the role of the equilibrium interfacial tension for the purposes of phase equilibria. This is however not a mechanical tension and indeed the interfaces between dense and dilute phases remain stable even when it is negative.  The reversal of Ostwald ripening by active processes is interesting, because in passive systems
Ostwald ripening is often difficult to arrest, let alone reverse. This is a main obstacle to the stabilisation of emulsions, where arrest can be achieved by trapping particles inside droplets that are insoluble in the surrounding fluid~\cite{higuchi1962physical,kabalnov1987ostwald,webster1998stabilization}. Ostwald ripening can however also be arrested by continuous chemical reactions~\cite{zwicker2017growth,Zwicker2015} or by continuous shearing~\cite{stansell2006nonequilibrium}, both of which maintain the system away from equilibrium, as do our active current terms.  

Our numerical studies confirm that the outcome of the reverse Ostwald process in AMB+ is a microphase separation, establishing this as a generic feature of active matter, without the neccessity of long-range interactions caused by hydrodynamics~\cite{tiribocchi2015active,matas2014hydrodynamic,thutupalli2017boundaries}, chemotaxis~\cite{liebchen2015clustering,saha2014clusters} or other system-specific causes~\cite{Brader:15,alarcon2017morphology,prymidis2015self,mani2015effect,mognetti2013living}.
The resulting microphase separated bubble and cluster phases should be seen as new (but intimately related) phases of active matter, whose detailed properties require further investigation in the future. We predict that each of these phases can either fill space in its own right, or coexist with a bulk phase comprising excess vapor (in the case of bubbles) or liquid (in the case of clusters).  In the former case, within the microphase-separated region, bubbles constantly form, migrate to the boundary, and pop into the vapor: we have called this `bubbly phase separation'. A similar phenomenon, with dense and dilute regions interchanged, is predicted at coexistence between a liquid cluster phase and excess liquid. Indeed, within AMB+ there is duality relation that maps every aspect of bubble behavior onto a corresponding one for clusters.

As stated in the introduction, cluster phases have been observed in experiments on active colloids~\cite{Palacci:12,Speck:13,thutupalli2017boundaries}, whereas bubbly phase separation has been observed in simulations of Active Brownian Particles with hard core repulsions~\cite{stenhammar2014phase}. This suggests a conjecture that, upon coarse graining to continuum level, these two systems map into opposite regions of the $(\lambda,\zeta$) parameter space of AMB+. So far, we are not aware of any observation -- either experimental or computational -- of either a homogeneous bubble phase, or of coexistence between a homogeneous dense phase and the cluster phase. We hope our prediction of their generic existence will motivate further investigations in this direction.

The statistical properties of the microphase-separated states are intriguing and also merit further study. In particular, we observed that the average bubble/cluster size increases when decreasing the noise level in the field theory. We so far have only a speculative explanation for this, involving a steady-state balance between deterministic and noisy terms in the evolution equations.


Also of interest is our finding of a transition line separating normal phase separation from bubbly phase separation. It would be good to identify specific microscopic models where this transition line can be studied; this might require independent control of the noise level and the mobility or interaction parameters. (There may hence be some active systems, perhaps including hard-core active Brownian particles, whose phase separation is always bubbly.) 
This transition signifies the presence of two distinct regimes, one where time reversal symmetry is almost restored at mesoscopic spatial and temporal scales, as speculated in the literature for over a decade~\cite{Speck2014PRL,Tailleur:08,fodor2016far,Brader:15,Maggi:15,szamel2016theory,nardini2017entropy}, and a second one where it remains strongly broken (in at least one of the coexisting phases). It would be natural to start such an investigation from the minimal model we were able to coarse-grain explicitly in Section \ref{explicit-coarse-graining}, where two-body forces are added to a quorum sensing model.

Meanwhile our coarse graining of that model confirms that nothing prevents the emergence from microscopics of the non-gradient currents governed by the activity parameter $\zeta$ at continuum level. These currents lead directly to reverse Ostwald ripening and microphase separation via a nonlocal contribution to the effective nonequilibrium chemical potential $\mu$ (as defined by $\nabla\cdot\bfJ = -\nabla^2\mu$). Our linking of microscopic to macroscopic models should help pave the way to future investigations of cluster phases and bubbly phase separation in terms of microscopic parameters, leading to better control of these nonequilibrium states of organization in simulations and, eventually, in experiments.

In this paper we have mainly worked at mean-field level and therefore not addressed the effects of activity near the liquid-vapor critical point, where fluctuations are dominant even in the passive limit. A preliminary numerical investigation of the critical dynamics of ABPs has recently appeared~\cite{siebert2017critical} suggesting, albeit inconclusively, that the universality class is altered from the passive case. A Renormalization Group study of AMB+ in the critical region, to be published elsewhere~\cite{caballeroBplus}, suggests that the combined action of $\zeta$ and $\lambda$ can indeed strongly change the behavior, but only beyond a finite activity threshold.

\section*{Acknowledgements} We thank F. Caballero, H. Chat\'e, R. Mari, T. Markovich, J. Stenhammar, J. Tailleur, and R. Wittkowski for discussions. 
CN acknowledges the support of an Aide Investissements d'Avenir du LabEx PALM (ANR-10-LABX-0039-PALM). Work funded in part
by the European Research Council under the Horizon
2020 Programme, ERC grant agreement number 740269. MEC is funded by the Royal Society.

{\em Author contributions.} ET and CN contributed equally to this work.

\appendix

\section{Direct numerical simulations} \label{simulations}
The dynamics of Active Model B+ is given by equations (\ref{eq:AMB+}-\ref{eq:AMB+J}).
In (\ref{eq:AMB+}), $\mathbf{\Lambda}(\bfr,t)$ is Gaussian white noise with zero mean and variance 
$\left<\Lambda_\alpha(\bfr,t)\Lambda_\beta(\bfr',t')\right>=\delta_{\alpha\beta}\delta(\bfr-\bfr')\delta(t-t')$
where Greek indices indicate spatial components.

All numerical simulations are performed at dimension $d=2$ in a square box of size $L\times L$ with periodic boundary conditions on all sides.
We choose $A=0.25$, $K=1$ and fix $K_1=0$ since this term does not change the observed phenomenology (see Appendix~\ref{K1}).
We discretise time as $t=n\Delta t$ and space as $x=i\Delta x$ and $y=j\Delta y$ where $n$, $i$, $j$ are integers.
We choose $\Delta t=0.001$ and $\Delta x=\Delta y=1$; our numerical simulations are converged (with respect to averaged quantities such as coarsening rates) at these values of $\Delta t$ and $\Delta x$. 

The dynamics of the field $\phi^n_{ij}$ then follows It\^{o} integration in time:
\qq
\phi^{n+1}_{ij} = \phi^{n}_{ij} - \Delta t  \nabla\cdot\mathbf{J}^n_{ij} - \sqrt{\frac{2D\Delta t}{\Delta x\Delta y}} \nabla\cdot\mathbf{\Gamma}^n_{ij} \label{eq:AMB+lattice}
\qqq
where $\Gamma^n_{\alpha ij}$ is now a Gaussian random variable with zero mean and variance 
$\left<\Gamma^m_{\alpha ij}\Gamma^n_{\beta kl}\right>=\delta_{mn}\delta_{\alpha\beta}\delta_{ik}\delta_{jl}$.
Now to compute the spatial derivatives, we use finite difference at order $\mathcal{O}(\Delta x^8)$:
\qq
\p_x\phi_{ij} &=& \frac{\frac{1}{280}\phi_{i-4,j} - \frac{4}{105}\phi_{i-3,j} + \frac{1}{5}\phi_{i-2,j} - \frac{4}{5}\phi_{i-1,j}  }{\Delta x} \nonumber\\
		&+&	 \frac{\frac{4}{5}\phi_{i+1,j} - \frac{1}{5}\phi_{i+2,j} + \frac{4}{105}\phi_{i+3,j} - \frac{1}{280}\phi_{i+4,j} }{\Delta x} \nonumber\\
		&+& \mathcal{O}\left(\Delta x^8\right) \label{eq:finite-difference}
\qqq 
and similarly for $\p_y\phi_{ij}$.
For second derivatives such as $\p_x^2\phi_{ij}$, we apply (\ref{eq:finite-difference}) twice~\cite{nardini2017entropy}.
This is to ensure that the equilibrium limit of (\ref{eq:AMB+lattice}) satisfies detailed balance exactly on the lattice.

\section{Binodals} \label{app:binodals}

We give details on the calculation of binodals $\phi_{1,2}$ for AMB+. 
It is useful for this to notice that the explicit solution to (\ref{eq:def-psi-g}) such that $\psi\to\phi$ and $g\to f$ in the passive limit ($\lambda=\zeta=0$)
is given by 
\qq 
\psi(\phi) &=& \frac{K}{\zeta-2\lambda} \left(e^{\frac{\zeta-2\lambda}{K} \phi} -1\right) \\
   g(\phi) &=& \frac{6 A K^4}{(\zeta-2\lambda) ^4}-\frac{A K^2}{(\zeta-2\lambda) ^2}-\frac{A K e^{\frac{(\zeta-2\lambda)  \phi }{K}}}{(\zeta-2\lambda) ^4} \\ \label{app:G-explicit}
                 && \Big[ -(\zeta-2\lambda) ^3 \phi  \left(\phi ^2-1\right)\nonumber\\
               &+& 6 K^3-6 \zeta  K^2 \phi +(\zeta-2\lambda) ^2 K \left(3 \phi ^2-1\right)\Big]\,.\nonumber
\qqq
Then, finding the binodals amounts to  find those $\phi_{1,2}$ such that the two conditions on bulk chemical potentials (\ref{eq:mu-flat}) and  on 
the pseudo-pressures (\ref{eq:pressure-flat2}) are satisfied. This is a simple numerical problem and the results are reported in Fig. (\ref{fig:binodals}).
As can be seen~\cite{Wittkowski14}, the binodals are at $\phi = \pm1$ only in the passive limit ($\zeta=\lambda=0$), which are the minima of the free energy density $f(\phi)$.

\section{Interface shape and perturbative corrections to binodals for curved interfaces} \label{perturbative}
We give here details about finite size corrections to binodals for spherically symmetric stationary solutions of dense droplets in a dilute environment in the mean-field limit ($D=0$), i.e. the computation of $\phi_{\pm}(R)$ as a function of $R$.

As emphasised in the main text,  the conditions (\ref{eq:mu-eq-curv}) and (\ref{eq:press-eq-curv2}) are not sufficient to find $\phi_{\pm}(R)$. Indeed they depend implicitly, via $\mathcal{S}_0,\mathcal{S}_1$, on the shape of the interface.
Here we discuss the structure of the interface of a spherical droplet, and 
show how to compute $w(\phi)=(\p_r\phi)^2$  perturbatively in $1/R$.
The knowledge of $w(\phi)$ allows us to compute 
$\mathcal{S}_0=\exp\left(\frac{\zeta-2\lambda}{K}\phi_+\right)\int_{\phi_+}^{\phi_-} \sqrt{w(x)} dx$ and 
$\mathcal{S}_1=\int_{\phi_+}^{\phi_-} \sqrt{w(x)}e^{\frac{\zeta-2\lambda}{K} x} dx$ (see equations (\ref{eq:S0}-\ref{eq:S1})). 
Combined with the stationary conditions (\ref{eq:mu-eq-curv}) and (\ref{eq:press-eq-curv2}), 
this fixes  $\phi_+(R)$ and $\phi_-(R)$ perturbatively to order $\mathcal{O}(1/R^2)$.

We consider a spherical droplet with density profile $\phi(r)$. As all along the paper we consider a dense droplet in a dilute environment.
In (\ref{eq:mu-total-curv}), we change variable from $\phi$ to $w$ \cite{Wittkowski14}, giving
\qq
w'(x) &=& \frac{2\lambda-\zeta}{K}w(x)  \nonumber \\
         &+& s(x,\phi_-,\mu_I(R)) + \mathcal{O}\left(\frac{1}{R^2}\right) \, \label{eq:app-interface-w-diff}
\qqq
where the prime now indicates derivative with respect to $x$ 
and we have defined:
\qq
s(x,\phi_-,\mu) &=&  \frac{2}{K}[f'(x) -\mu_I(R)] - \frac{2(d-1)}{R}\sqrt{w(x)} \nonumber\\
				&+&\frac{2(d-1)}{K R} \zeta \int_x^{\phi_-} \sqrt{w(y)} \,dy
\,.
\qqq
If we temporarily ignore the functional dependence of $s$ on $w$, and assume that $\mu_I(R)$ is known, 
Eq. (\ref{eq:app-interface-w-diff}) is linear and  can be solved as
\qq\label{eq:app-interface-w-sol-gen}
w(x)&=&e^{-\frac{\zeta-2\lambda}{K} x} \\
&&\left[ C + \int_1^x e^{\frac{\zeta-2\lambda}{K} y } s(y,\phi_-,\mu_I(R)) dy \right] + \mathcal{O}\left(\frac{1}{R^2}\right) \,.\nonumber
\qqq
The knowledge of $\phi_+,\phi_-$ then allows us to fix the integration constant $C$.

The perturbative strategy then goes as follows.
At $\mathcal{O}(1/R^0)$, we can approximate $\phi_+$ and $\phi_-$ to be the binodals, \emph{i.e.} $\phi_{\pm}\simeq\phi_{1,2}$ and, at this level of approximation,  the chemical potential is also known $\mu_I(R)=\bar\mu$ .
We can thus obtain $w(\phi)$ from (\ref{eq:app-interface-w-sol-gen}).
Using this value of $w(\phi)$, we can then obtain $\mathcal{S}_0$ and $\mathcal{S}_1$.
Once these are known, 
we can solve the simultaneous equations (\ref{eq:mu-eq-curv}) and (\ref{eq:press-eq-curv2}) to obtain $\phi_{\pm}$, along with $\mu_I(R)$ at next order $\mathcal{O}(1/R)$.
Using these new values of $\phi_{\pm}$ we again calculate $w(\phi)$, $\mathcal{S}_0$ and $\mathcal{S}_1$, and 
again solve (\ref{eq:mu-eq-curv}) and (\ref{eq:press-eq-curv2}) to get $\phi_{\pm}$ at $\mathcal{O}(1/R^2)$.
We cannot go further because (\ref{eq:app-interface-w-diff}) is obtained discarding terms of order $\mathcal{O}\left(1/R^2\right)$ which would otherwise make the equation for $w$ non-autonomous. However, we already have all the results we need.
This perturbative solution was implemented numerically to give the results in Fig.~\ref{fig:phi+-}.

\section{Effect of $K_1\neq 0$} \label{K1}
As stated in the main text, setting $K_1\neq 0$ does not affect qualitatively any of the conclusions of our work. All our results can be explicitly generalized to this case, but we restrict ourselves here to a few statements about how $K_1$ enters the calculations.

First of all, the chemical potential in spherical symmetry now reads
\qq\label{eq:stationary-AMB+explicit-lambda-zeta}
\mu
&=&
f'(\phi) -K(\phi) \phi'' -\frac{(d-1)K(\phi)}{r} \phi' \nonumber
\\&+&\nu
\phi'^2
+(d-1)\zeta \int_r^\infty \frac{\phi'^2(y)}{y} \,dy\,
\qqq
where we recall that $K(\phi)=K+2K_1\phi$ and we have defined $\nu=\Big(\lambda-K_1-\frac{\zeta}{2}\Big)$.

Let us first consider the calculation of the binodals $\phi_{1,2}$. This goes similarly to the one presented in Section~\ref{binodals} except that $\psi$ takes the form
\qq\label{eq:pseudo-press-h}
\psi_{K_1}(\phi) &=& - \frac{K}{2\nu}\left(\bar{\psi}(\phi)-1\right)\\
 \bar{\psi}(\phi)&=&\left[ 1+\frac{2K_1\phi}{K} \right]^{-\frac{\nu}{K_1}} 
\qqq
while $g_{K_1}$ is the solution to $\p g_{K_1}/\p \psi_{K_1} = \p f/\p\phi$ that we do not report here explicitly. 
For $K_1\neq 0$, $\phi_{1,2}$ can thus be obtained solving (\ref{eq:mu-flat}), (\ref{eq:pressure-flat2}) with $\psi$ and $g$ replaced by $\psi_{K_1}$ and $g_{K_1}$.

Similarly, when considering a droplet of radius $R$, the values of $\phi_{\pm}(R)$ are also affected by $K_1\neq 0$. 
Again, the calculations are similar to those in Section~\ref{binodals} and \ref{Ostwald-ripening} but with $\psi$ and $g$ replaced by $\psi_{K_1}$ and $g_{K_1}$.

Generalising the argument leading to Ostwald ripening presented in Section~\ref{Ostwald-ripening} to the case of $K_1\neq 0$, 
we find that (\ref{eq:R-time-evolution})-(\ref{eq:beta}) still hold after replacing $\sigma$ with
\qq\label{eq:pseudo-sigma-K1}
\sigma_{K_1} = -\frac{K}{2\nu} \left[ \zeta \mathcal{S}_2-( 2\lambda-2K_1)\mathcal{S}_3 \right]
\qqq
where $\mathcal{S}_2,\mathcal{S}_3$ are parameters that depends on the shape of the interface and are defined by
\qq
\mathcal{S}_2&\equiv& \bar{\psi}(\phi_2) \int_0^\infty \phi'^2(y) \,dy\\
\mathcal{S}_3&\equiv& \int_{0}^{\infty} \bar{\psi}(\phi(y))\, \phi'^2(y) dy\,.
\qqq

Conditions on $\lambda,\zeta,K_1$ such that $\sigma_{K_1}<0$ can be obtained from (\ref{eq:pseudo-sigma-K1}), leading again to reversed Ostwald ripening.  We leave the detailed study of the phase diagram in terms of $\lambda,\zeta$, and $K_1$ to future studies.

\section{Evolution of two droplets} \label{app:two-droplets}
In this Appendix, we adapt the argument of Section \ref{Ostwald-ripening} to explicitly deal with the initial condition with two droplets as in Fig.~\ref{fig:phase-diagram-MF}(b), and predict the full time evolution of their radii $R_{1,2}(t)$. We believe these results to be interesting by themselves, as we have found nowhere in the literature -- not even the equilibrium literature -- an adaptation of the argument leading to Ostwald ripening able to track the evolution of two droplets. Moreover, it will turn out clearly that the quantity $\epsilon$ used in Section \ref{Ostwald-ripening} has to be interpreted as the supersaturation among distant droplets, and it is thus negative when $\sigma<0$. This is a subtle but important point as it implies, as already stressed in Section~\ref{Ostwald-ripening}, that $\epsilon<0$ when $\sigma<0$.

First we consider droplet $1$ whose radius is $R_1(t)$. We assume it to be stationary, and thus its density inside to be $\phi_+(R_1)$, while the density just outside has to be $\phi_-(R_1)$.
We furthermore assume that, at infinity, droplet $1$ ``sees'' the outer density of droplet $2$, \emph{i.e.} $\phi_-(R_2)$.
Following the derivation in Section~\ref{Ostwald-ripening}, but without using the approximation $\phi_{\pm}\simeq \phi_{1,2}$,
the time evolution of $R_1(t)$ is then (for dimension $d=3$): 
\qq\label{eq:R1dot}
\Delta\phi(R_1)\,\dot{R}_1 = \frac{f'(\phi_-(R_2)) - f'(\phi_-(R_1))}{R_1 } \,,
\qqq
where $\Delta\phi(R_1)=(\phi_+(R_1)-\phi_-(R_1))$; 
we recognise $f'(\phi_-(R_1))$ to be the chemical potential just outside the interface of droplet $1$ and $f'(\phi_-(R_2))$ to be the chemical potential which droplet $1$ ``sees'' at infinity.
Note that in $d=2$, there will be a logarithmic correction but this does not change the phenomenology.
Similarly, we assume droplet $2$ to be stationary, and thus to be formed by the densities $\phi_{\pm}(R_2)$, but to ``see'' the outer density of droplet $1$ at infinity.
The time evolution of $R_2(t)$ is then:
\qq\label{eq:R2dot}
\Delta\phi(R_2)\, \dot{R}_2 = \frac{f'(\phi_-(R_1)) - f'(\phi_-(R_2))}{R_2} \,.
\qqq

It is now important to recall that we were actually able to compute $\phi_{\pm}(R)$ in Section \ref{binodals}, see Fig.~\ref{fig:phi+-}.
Using these values, we can solve (\ref{eq:R1dot},\ref{eq:R2dot}) simultaneously to obtain $R_1(t)$ and $R_2(t)$.
As stressed in the main text, these predictions where compared to direct numerical simulation of AMB+ in Fig.~\ref{fig:phase-diagram-MF}(b), finding good qualitative agreement. Quantitative agreement is not expected, since the calculation leading to (\ref{eq:R1dot},\ref{eq:R2dot}) 
assumes an effective medium picture, whereby the presence of the second droplet breaks the spherical symmetry in the neighborhood of the first droplet. Moreover, it was assumed that the two droplets are separated by a distance $\ell\gg R_{1,2}$, a condition which is not really met in simulations due to computational limitations. Finally, for a precise quantitative comparison, one shall take care of the fact that simulations are performed in $d=2$ with periodic boundary conditions.

\section{Current flowing between two droplets} \label{app:currents}

\begin{figure}
\begin{centering}
\includegraphics[width=0.9\columnwidth]{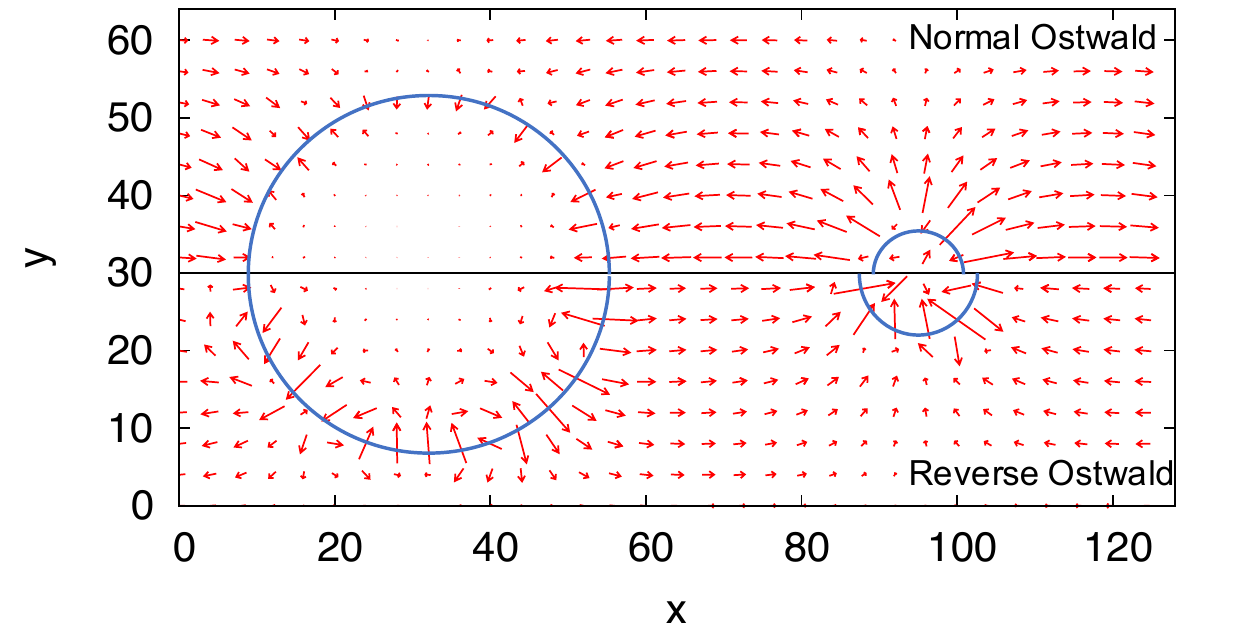}
\par\end{centering}
\caption{ 
Snapshots of the total current $\mathbf{J}$ for the two droplet simulations shown in Fig.~\ref{fig:phase-diagram-MF}(b) at short time $t=1000$:
top half corresponds to regime A or normal Ostwald ($\zeta=-0.8$, $\lambda=-0.4$) and bottom half corresponds to regime B or reverse Ostwald ($\zeta=-1.6$, $\lambda=-0.8$).
(Note that the system is symmetric in $y$. The arrows are magnified for better visibility, typical magnitude of $\mathbf{J}$ is around $10^{-4}$.)} 
\label{fig:current}
\end{figure}

Fig.~\ref{fig:current} shows the total current $\mathbf{J}$ for the two droplet/cluster state shown in Fig.~\ref{fig:phase-diagram-MF}(b) at short time $t=1000$.
For regime A in the mean field phase diagram, current flows from small droplet to big droplet (Fig.~\ref{fig:current} top)
and for regime B in the mean field phase diagram, current flows from big droplet to small droplet (Fig.~\ref{fig:current} bottom) as expected.

\section{Details on explicit coarse-graining}\label{app:cg}
In this Appendix, we give details about the coarse-graining procedure developed in Section \ref{explicit-coarse-graining} and explicitly
connect microscopic parameters of the particle model (\ref{eq:AOUPs-main-1},\ref{eq:AOUPs-main-2}) to those entering in its hydrodynamic description.

\subsection{Contribution due to quorum sensing }

We first consider the quorum sensing part and give the explicit expressions of $f_{qs}, K^{qs}, K_1^{qs}$ and $\lambda_{qs}$ in terms of microscopic parameters. 

We first expand $v[\rho]$ at lowest order in gradients and retain only terms allowed by rotational symmetry:
$ v[\rho]=v(\rho)+\beta(\rho) |\nabla \rho|^2 +\gamma(\rho) \nabla^2\rho$. Moreover, we Taylor expand $v(\rho),\beta(\rho),\gamma(\rho)$ with the notation
\qq
v(\rho)&=& v_0+v_1\rho  +o(\rho)\nonumber\\
\gamma(\rho)&=&\gamma_0+\gamma_1\rho + o(\rho)\label{app:eq:Vphi-gammaphi-beta-phi}\\
\beta(\rho)&=&\beta_0 + o(\rho^0)\nonumber
\qqq
where higher order terms are irrelevant at the order considered. From (\ref{eq:Dean-Many-AMB+muqs}) and (\ref{app:eq:Vphi-gammaphi-beta-phi}), we obtain
\qq\label{app:eq:mu-qs-0}
\mu_{qs} &=& 
f'_{qs}(\rho)+ \frac{v(\rho)}{\tau v^2(\rho)+T}\times\\
&&\times\left[\beta(\rho) |\nabla \rho|^2 +\gamma(\rho) \nabla^2\rho\right]
+o(\nabla^2\rho^2)\,.\nonumber
\qqq
Expanding in powers of $\rho$ leads to (\ref{app:eq:cg-qs-complete},\ref{eq:Fqs-prx}), where the local part of the free energy is
\qq\label{app:-prx-fqs}
f_{qs} = \,\rho(\log\rho-1) +\frac{1}{2}\int^\rho d\varphi\, \log(T+\tau v^2(\varphi))\,
\qqq
and the other parameters are set by 
\qq
K^{qs}&=&-\frac{v_0\gamma_0}{2(T+\tau v_0^2)}\\
K_1^{qs}&=& -\frac{1}{2(T+\tau v_0^2)}
\left[
v_0\gamma_1+v_1\gamma_0-\frac{2v_0^2v_1\gamma_0\tau}{T+\tau v_0^2}
\right]\\
\lambda_{qs}&=& \frac{v_0\beta_0}{T+\tau v_0^2}+K_1^{qs}\,.\label{app:prx-eq:cg-qs-complete-1}
\qqq

\subsection{Contribution due to the two-body potential $U$}
We finally compute the effect of $U$ to get the form of $\mu_r$ in (\ref{eq:cg-Jtilde}) and derive (\ref{eq:zeta-explicit}-\ref{app:eq-BC-cg-2}). 
We first expand the convolution $U\star \rho$ in gradients of the density
\qq\label{app:eq:non-grad-J-Taylor-AB}
\nabla_\alpha (U\star \rho) = C \nabla_\alpha \rho + B\nabla_\alpha \nabla^2\rho +o(\nabla^3)
\qqq
where $B,C$ are given in (\ref{app:eq-BC-cg-2},\ref{app:eq-BC-cg-1}). It is then easy to see that (\ref{eq:cg-Jtilde}) holds, with $\zeta_{cg}$ given in (\ref{eq:zeta-explicit}) and 
\qq\label{eq:mu-residual}
\mu_r = \frac{\delta \mathcal{F}_r}{\delta \rho}+\lambda_r |\nabla\rho|^2\,.
\qqq
where we have defined
\qq\label{app:eq:mur}
\mathcal{F}_r&=&\int  f_r(\rho) \,+\frac{K^r+2K_1^r\rho}{2}|\nabla\rho|^2\,d\bfr\\
f_r(\rho)&=&\int^{\rho}d\phi
\int^\phi d\psi \frac{C}{T+\tau v^2(\psi)}
\qqq
with
\qq
&K^r= -\frac{B}{T+\tau v_0^2}\\
&K_{1}^r =\lambda_r=\frac{B\tau v_0v_1}{(T+\tau v_0^2)^2}\,.\label{eq:mu-residual-parameters-K1-lambda}
\qqq

\bibliography{biblio}
\end{document}